\documentclass[twocolumn,twocolappendix]{aastex631}
\usepackage{amsmath,amssymb}
\usepackage{multirow}
\usepackage{lipsum}

\shorttitle{Insufficiency of Homogeneous Gyrosynchrotron Model for CMEs}
\shortauthors{Kansabanik et al.}

\begin{document}

\title{Spectropolarimetric Radio Imaging of Faint Gyrosynchrotron Emission from a CME : A Possible Indication of the Insufficiency of Homogeneous Models}

\author[0000-0001-8801-9635]{Devojyoti Kansabanik}
\affiliation{The Johns Hopkins University Applied Physics Laboratory, 11101 Johns Hopkins Road, Laurel, MD 20723, USA}
\affiliation{Cooperative Programs for the Advancement of Earth System Science, University Corporation for Atmospheric Research, Boulder, CO, USA}
\affiliation{National Centre for Radio Astrophysics, Tata Institute of Fundamental Research, S. P. Pune University Campus, Pune 411007, India}

\author[0000-0002-2325-5298]{Surajit Mondal}
\affiliation{Center for Solar-Terrestrial Research, New Jersey Institute of Technology, 323 M L King Jr Boulevard, Newark, NJ 07102-1982, USA}

\author[0000-0002-4768-9058]{Divya Oberoi}
\affiliation{National Centre for Radio Astrophysics, Tata Institute of Fundamental Research, S. P. Pune University Campus, Pune 411007, India}

\correspondingauthor{Devojyoti Kansabanik}
\email{dkansabanik@ucar.edu, devojyoti96@gmail.com}

\begin{abstract}
The geo-effectiveness of coronal mass ejections (CMEs) is determined primarily by their magnetic fields. Modeling of Gyrosynchrotron (GS) emission is a promising remote sensing technique to measure the CME magnetic field at coronal heights. However, faint GS emission from CME flux ropes is hard to detect in the presence of bright solar emission from the solar corona. With high dynamic-range spectropolarimetric meter wavelength solar images provided by the Murchison Widefield Array, we have detected faint GS emission from a CME out to $\sim 8.3\ R_\odot$, the largest heliocentric distance reported to date. High-fidelity polarimetric calibration also allowed us to robustly detect circularly polarized emission from GS emission. For the first time in literature, Stokes V detection has jointly been used with Stokes I spectra to constrain GS models. One expects that the inclusion of polarimetric measurement will provide tighter constraints on GS model parameters. Instead, we found that homogeneous GS models, which have been used in all prior works, are unable to model both the total intensity and circular polarized emission simultaneously. This strongly suggests the need for using inhomogeneous GS models to robustly estimate the CME magnetic field and plasma parameters.     
\end{abstract}


\section{Introduction}
\label{sec:intro}
Large-scale eruptions of plasma and magnetic fields from the solar corona into the heliosphere are known as Coronal Mass Ejections (CMEs). CMEs are the most crucial phenomena determining the space weather. The geo-effectiveness of a CME is determined primarily by its magnetic field strength and topology \citep{Vourlidas2019,Ionescu2021}. CME magnetic fields may get modified as they propagate through the corona and the heliosphere due to interactions with the coronal magnetic fields, interplanetary magnetic fields, and other heliospheric structures. Observations at radio wavelengths provide a few different useful methods to measure the plasma parameters of the CMEs at coronal and heliospheric heights. These include thermal free-free emission from CME plasma \citep[e.g.,][etc.]{Gopalswamy1992,Gopalswamy1993,Ramesh_2003,Ramesh2021}, coherent plasma emissions from CME shocks (type-II radio bursts) \citep[e.g.,][etc.]{Nelson1985,Gopalswamy_2000_typeII,Cairns2003,Gopalswamy2019_typeII,Jebaraj2021}, coherent plasma emissions from CME cores (type-IV radio bursts) \citep[e.g.,][etc.]{KRISHNAN1961,Ramaty1969_typeIV,Kumari2017_typeIV,morosan2019} and gyrosynchrotron (GS) emission from CME plasma \citep[e.g.,][etc.]{bastian2001,Tun2013,Bain2014,Carley2017,Mondal2020a}. At coronal heights GS emission from mildly relativistic electrons gyrating in the CME magnetic field \citep[e.g.,][etc.]{bastian2001,Tun2013,Mondal2020a} is one of these few methods that can be used to measure the CME-entrained magnetic fields. Since the first imaging detection and modeling by \cite{bastian2001}, there have been only a handful of studies that have successfully managed to detect faint GS emission from CME loops \citep{Maia2007,Carley2017,Mondal2020a}. This scenario is changing over the past few years with the high dynamic-range (DR) and high-fidelity spectropolarimetric meter wavelength solar radio images provided a robust polarimetric calibration and imaging algorithm -- {\it Polarimetry using Automated Imaging Routine for the Compact Arrays for the Radio Sun} \citep[P-AIRCARS;][]{Mondal2019,Kansabanik2022,Kansabanik2022_paircarsI,Kansabanik_paircars_2} using observations with the Murchison Widefield Array \citep[MWA;][]{Tingay2013,Wayth2018}. It is now possible to detect much fainter GS emissions from CMEs with MWA solar observations \citep{Kansabanik2023_CME1}.

Even with the routine and reliable detection of GS emission from CMEs, estimating the plasma parameters from the observed GS spectrum remains challenging. The GS model requires ten free parameters even for the simplest homogeneous and isotropic plasma distributions with a single power-law energy distribution of non-thermal electrons \citep{Fleishman_2010,Kuznetsov_2021}. Constraining all of these GS model parameters only using the total intensity (Stokes I) spectrum is not possible and requires several assumptions to be made. Using the high-fidelity and high DR spectropolarimetric imaging with the MWA, K23 demonstrated that the availability of even strong upper limits on the Stokes V measurements, along with the Stokes I spectra, can significantly improve the constraints on the GS model parameters and lift some of the degeneracies in the model parameters. We note that this, and all prior work on modeling of CME GS spectra, assume homogeneous distributions of all plasma parameters along the line-of-sight (LoS).

The radio emission from the CME under consideration here is significantly brighter than the one studied by K23 and this has helped us detect the Stokes V emission from a part of the CME contrary to K23, where authors were only able to provide an upper limit to the absolute Stokes V flux density. For the first time, this work simultaneously uses constraints from a Stokes V detection along with Stokes V upper limits and Stokes I detections to constrain the GS model parameters.

This article is organized as follows --Section \ref{sec:obs} describes the observation and the data analysis. The imaging results are presented in Section \ref{sec:results}, along with the discussion about the origin of the radio emission. Section \ref{sec:spectra_model} describes spectral modeling using a homogeneous GS model. Validity of the homogeneous and isotropic assumptions of GS emission to model the observed spectra is discussed in Section \ref{subsec:homo_insuff} followed by simulations to explore the effects of inhomogeneity of plasma parameters along the LoS in Section \ref{sec:nonuniform}. Section \ref{sec:conclusion} presents a conclusion of the work followed by a discussion of future work in Section \ref{sec:future_work}.

\begin{figure}
    \centering
     \includegraphics[trim={0cm 0cm 2cm 0cm},clip,scale=0.45]{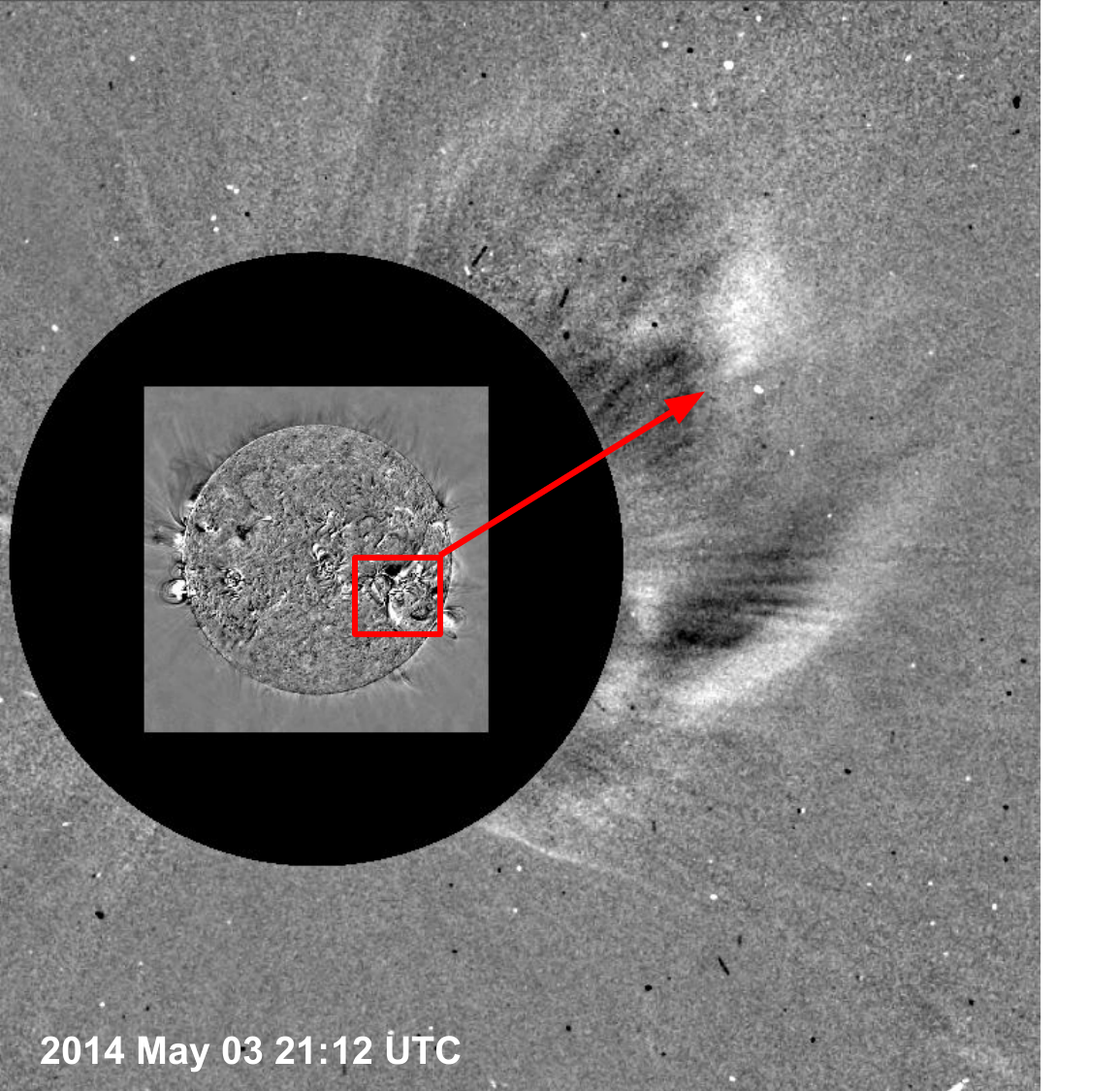} 
     \caption{\textbf{Eruption of CME-2 observed using SDO/AIA spacecraft.} Eruption of CME-2 observed using SDO/AIA spacecraft. CME-2 erupted from the visible part of the solar disc. A composite base difference image from the Atmospheric Imaging Assembly (AIA) onboard SDO at 171 \AA\ and LASCO C2 coronagraph image onboard the SOHO spacecraft is shown. The red box shows the active region 12047, which is the eruption site for CME-2, and the red arrow shows the propagation direction.}
    \label{fig:south_cme_eruption}
\end{figure}

\section{Observation and Data Analysis}\label{sec:obs}
We present the MWA observation on 04 May 2014 in this study. A total of six active regions were present on the visible part of the solar disk\footnote{\href{https://www.solarmonitor.org/?date=20140504}{Flare list on solarmonitor.org}} on this day. Although no large flares (M or X GOES class) are reported, a total of nine CMEs erupted on this day as reported in the CME catalog provided by the Coordinated Data Analysis Workshop (CDAW)\footnote{\href{https://cdaw.gsfc.nasa.gov/CME_list/UNIVERSAL/2014_05/univ2014_05.html}{CDAW CME list on 04 May 2014}}. Most of these CMEs are reported as ``poor events". Of these, two CMEs overlap with the MWA observing window -- one of them propagates towards solar north (CME-1) and the other towards southwest (CME-2). This paper focuses on CME-2. A detailed spectropolarimetric imaging and modeling study of the CME-1 was presented in \cite{Kansabanik2023_CME1} (hereafter K23).

\subsection{Eruption and Evolution of CME-2}\label{subsec:evolution}
The CME-2 erupted from the active region 12047 present on the visible part of the Sun. The eruption site is the bright spot inside the red box in Figure \ref{fig:south_cme_eruption}. CME-2 first appeared in the field-of-view (FoV) of the C2 coronagraph of the Large Angle Spectroscopic Coronagraph \citep[LASCO;][]{Brueckner1995} onboard the Solar and Heliospheric Observatory \citep[SOHO;][]{Domingo1995} at 20:48 UTC on 03 May 2014. It was visible in LASCO C3 coronagraph until 02:06 UTC on 2014 May 04 up to about 17 $R_\odot$. The CDAW catalog classified the CME-2 as a partial halo CME.

\subsection{Radio Observation and Data Analysis}\label{subsec:radio_analysis}
CME-2 was observed at meter-wavelength radio bands using the MWA on 2014 May 04 from 00:48 UTC to 07:32 UTC under the project ID  G0002\footnote{\href{http://ws.mwatelescope.org/metadata/find}{MWA Archive}}. The MWA observing band is split into 12 frequency bands, each of width 2.56 MHz, and centered around 80, 89, 98, 108, 120, 132, 145, 161, 179, 196, 217, and 240 MHz. The temporal and spectral resolution of the MWA data were 0.5 $\mathrm{s}$ and 40 $\mathrm{kHz}$, respectively. Typically different types of radio bursts, such as type-II, -III, and/or -IV \citep{Gopalswamy2011_CME_radio,Carley2020}, are associated with CMEs. During this time no solar radio bursts were reported \citep{Kansabanik2023_CME1}. We performed the polarization calibration and full Stokes imaging of the MWA observations using P-AIRCARS. Integration of 10 s and 2.56 MHz was used for imaging for all 12 frequency bands. All polarization images follow the IAU/IEEE convention of Stokes parameters \citep{IAU_1973,Hamaker1996_3}. 
\begin{figure}[!htbp]
    \centering
    \includegraphics[scale=0.43]{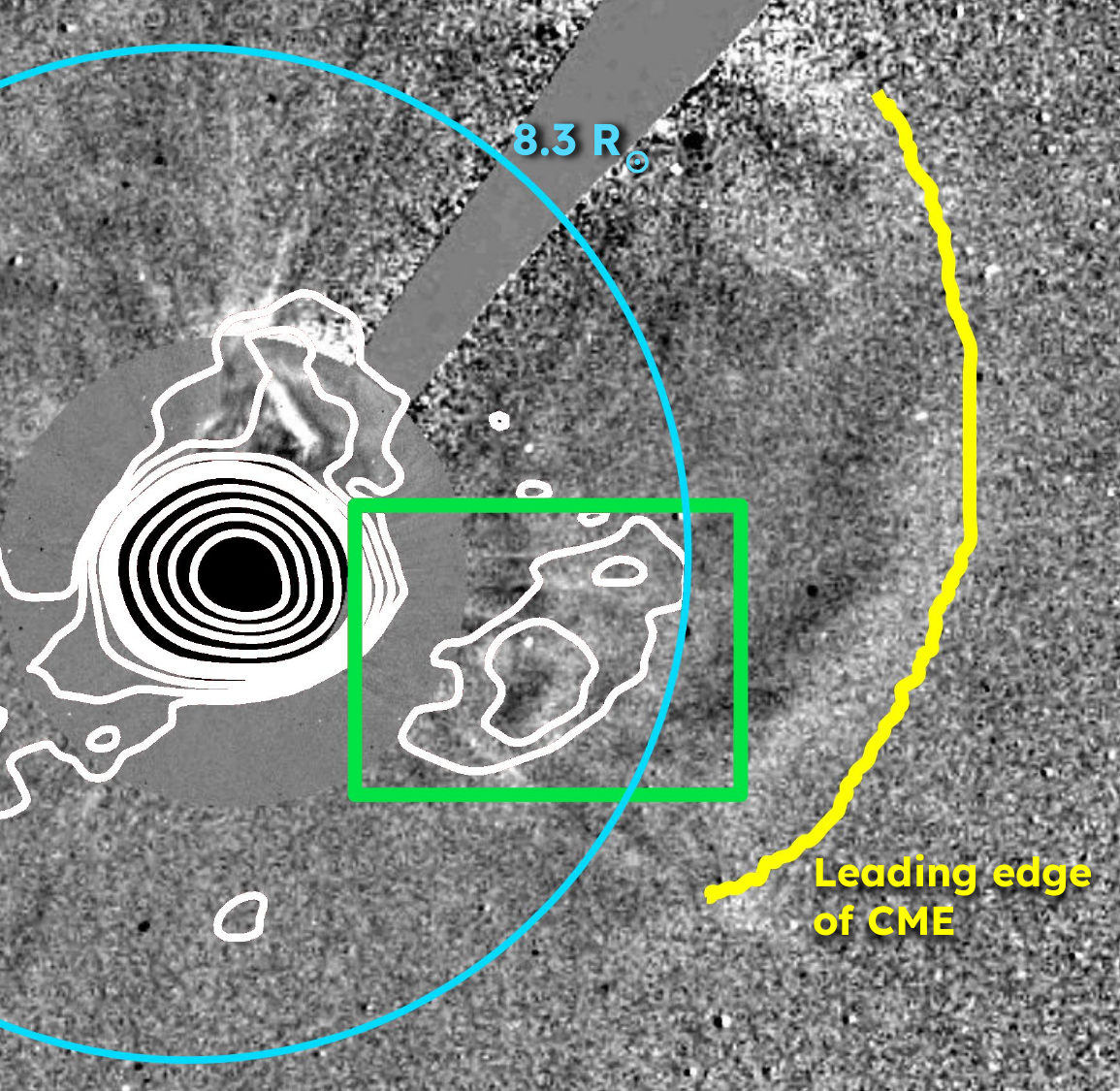}
    \caption{Stokes I emissions at 80 MHz are shown by the white contours overlaid on the LASCO C2 and C3 running difference image of CME-2 at 01:30 UTC on 2014 May 04. The radio image is at 01:24:55 UTC. Contour levels are at  0.5, 1, 2, 4, 6, 8, 20, 40, 60, and 80 \% of the peak flux density. Radio emission marked by the green box is associated with CME-2 and is the focus of this work. The emission is detected up to 8.3 $R_\odot$, shown by the cyan circle. The yellow curved line shows the leading edge of the CME-2.}
    \label{fig:south_cme}
\end{figure}

\section{Results}\label{sec:results}
This section presents the results from the wideband spectropolarimetric imaging observation of CME-2 using the MWA and the possible mechanisms that can give rise to it.

\subsection{Radio Emission from CME-2}\label{subsec:radio_emission_CME2}
Figure \ref{fig:south_cme} shows a sample Stokes I image at 80.62 MHz using the white contours overlaid on LASCO C2 and C3 running difference images. This work focuses on the radio emission associated with CME-2 marked by the green box. The study of another extended radio emission feature seen towards the southeast in Figure \ref{fig:south_cme} is co-located with a streamer and is beyond the scope of this work. 

\begin{figure*}[!htbp]
    \centering
    \includegraphics[trim={0cm 0cm 0cm 0cm},clip,scale=0.28]{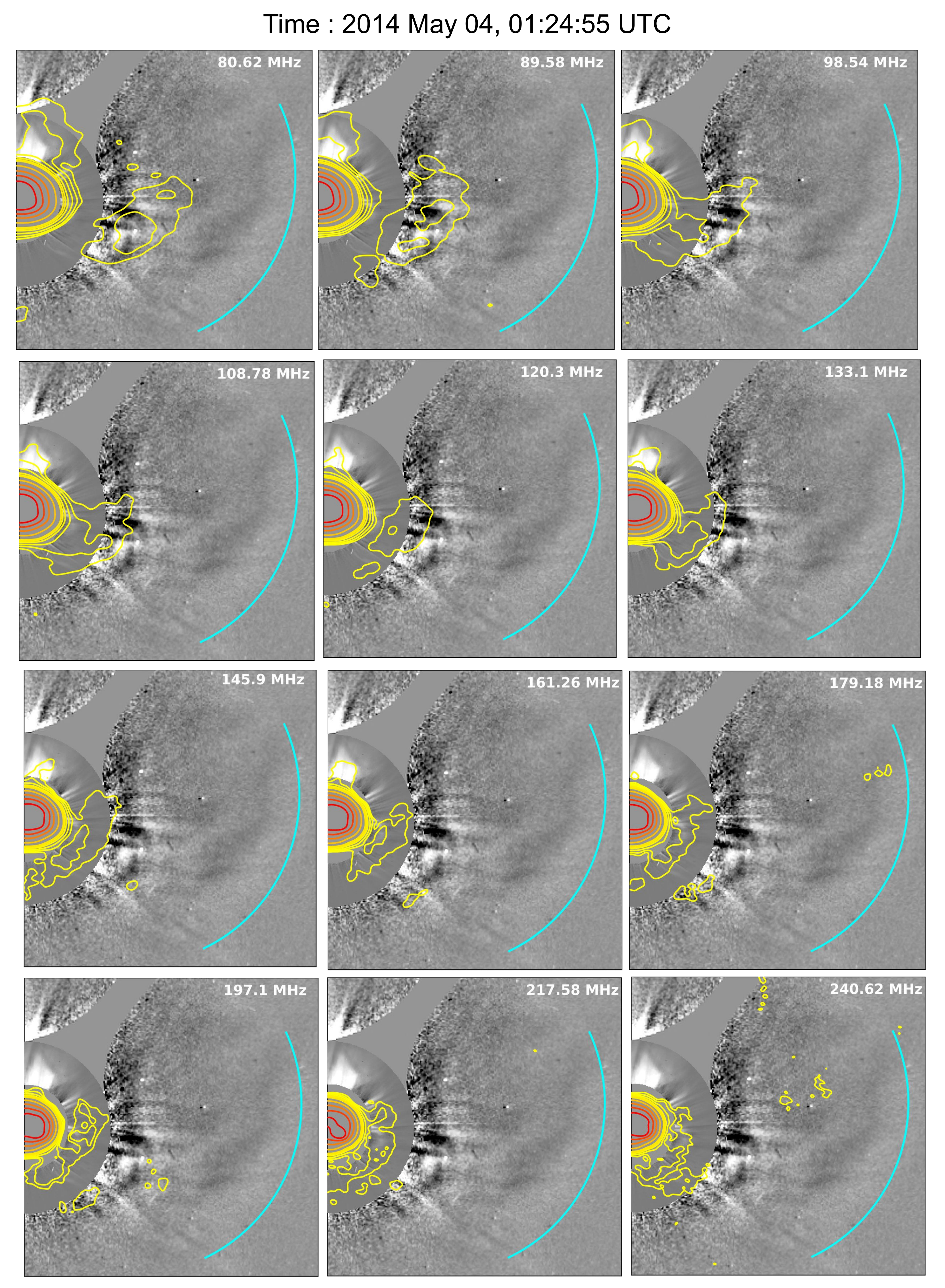}
    \caption{\textbf{Stokes I radio emission from CME-2 at MWA frequency bands at 2014 May 04, 01:24:55 UTC.} Stokes I emissions are shown by contours overlaid on LASCO C2 and C3 base difference images. Frequency increases from the top left panel of the image to the bottom right panel. Contour levels are at 0.5, 1, 2, 4, 6, 8, 20, 40, 60, and 80 \% of the peak flux density. Cyan arcs mark the faint leading edge of the CME-2.}
    \label{fig:c2_c3_comp_freq}
\end{figure*}

The extended radio emission is detected in all 12 frequency bands each of 2.56 MHz bandwidth, centered at 80, 89, 98, 108, 120, 132, 145, 161, 179, 196, 217, and 240 MHz. The evolution of the radio emission with frequency for a single time slice centered at 01:24:55 UTC is shown in Figure \ref{fig:c2_c3_comp_freq}. Frequency increases from the top left to the bottom right of the figure. The spatial extent of radio emission shrinks to lower heliocentric heights with increasing frequency. At the lowest frequency, 80 MHz, the radio emission extends up to 8.3 $R_\odot$, while at 240 MHz the emission is seen only out to $\sim2\ R_\odot$. 
\begin{figure*}
\begin{interactive}{animation}{CME2_movie.mp4}
\centering
\includegraphics[trim={0cm 0cm 0cm 0cm},clip,scale=0.145]{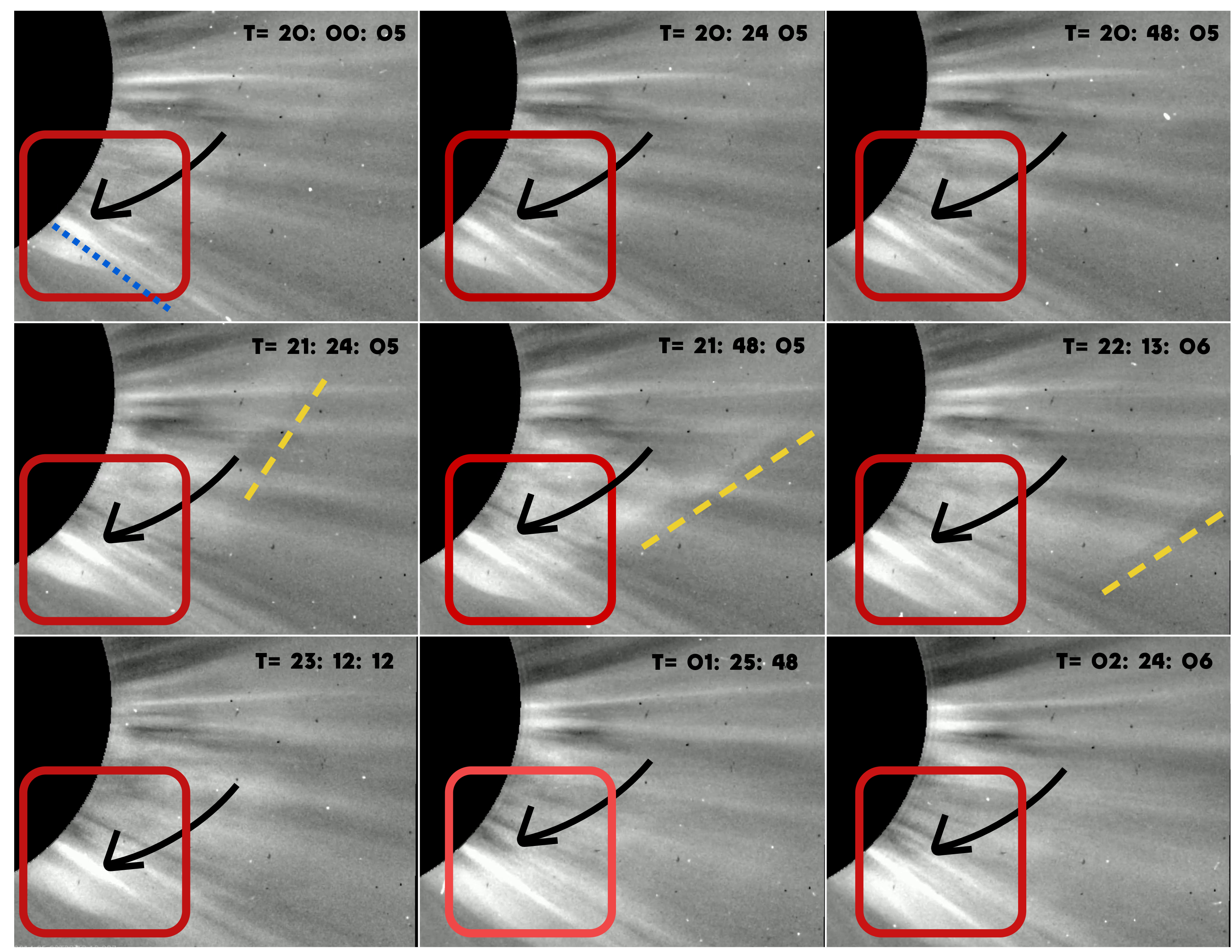}
\end{interactive}
\caption{Disturbances in magnetic fields are seen in the movie as the CME-2 passes over the region marked by red boxes in the static figure. The movie starts on 03 May 2014 15:12:08 UTC, and ends on 04 May 2014 02:36:05 UTC, covering the first appearance of CME-2 in the LASCO C2 FoV at 03 May 2014 20:48 UTC. The static figure shows some snapshots of the movie with increasing time from the top-left to the bottom-right corner. The first row shows three snapshots before CME-2 appeared in the LASCO C2 FoV on 03 May 2014, at 20:00:05, 20:24:04, and 20:48:05 UTC, respectively. In these three frames, the streamer marked by black arrows maintains its morphology and shows two distinct sharp features. The region between these two distinct features is marked by a blue dotted line in the first panel. In the second row, three snapshots are shown while CME-2 is passing the streamer region, at 03 May 2014, 21:24:05, 21:48:05, and 22:13:06 UTC. The faint leading edge of the CME is marked by yellow dashed lines. During this time the streamer starts changing its morphology, becoming wider and more diffused. The third row shows snapshots after CME-2 has passed the streamer region and at 23:12:12 on 03 May 2014 and 01:25:48, and 02:24:06 UTC on 04 May 2014, respectively. In these three snapshots, the two distinct components of the streamer almost merge and make a much wider diffused structure. These signatures show disturbances in the magnetic field structures of the streamer region after the CME-2 has passed over there. (A base difference movie of LASCO C2 is available here. Timestamps are at the bottom left corner of the movie.)}
\label{movie:CME2}
\end{figure*}

\subsection{Possible Radio Emission Mechanisms}
\label{subsec:origin_of_emission}
Metre-wavelength radio emission associated with CMEs can arise due to several possible emission mechanisms -- plasma emission, thermal free-free emission, and GS emission. Hence, before modeling the observed spectra, one needs to identify the dominant emission mechanism responsible for the emission. To determine the emission mechanism of the observed radio emission from CME-2, we have followed a similar path as described in \cite{Kansabanik2023_CME1}.

First, we looked for the possibility of plasma emission. For that, the average coronal electron density ($n_\mathrm{e}$) is estimated from the LASCO-C2 white light coronagraph images, using the inversion method developed by \cite{Hayes2001}. Estimated $n_\mathrm{e}$ varies from $\sim0.7\times10^6\ \mathrm{cm^{-3}}$ to $\sim1.25\times10^6\ \mathrm{cm^{-3}}$ at the location of the radio emission associated with CME-2. This leads to a corresponding plasma frequency of about $7.1\ \mathrm{MHz}$, more than an order of magnitude lower than the frequencies (a few hundred MHz) at which radio emission from CME-2 is detected. This rules out plasma emissions as a possible mechanism. 

The next possibility that we examined is the thermal free-free emission. The emission could arise from either optically thick or thin free-free emission. We have estimated free-free optical depth as \citep{Gary1994},
\begin{equation}
    \tau_\nu\approx0.2\ \frac{\int n_e^2\ dl}{\nu^2\ T_e^{\frac{3}{2}}}.
    \label{eq:free_free_optical_depth}
\end{equation}
where, $T_\mathrm{e}\approx10^6\ \mathrm{K}$ is the coronal plasma temperature, $\nu$ is the emission frequency and $l$ is the length of the path segment along the LoS. In this expression, magnetic field contribution is not considered, because that does not affect Stokes I emission.

The electron density drops rapidly with increasing coronal height, dropping by more than an order of magnitude between 2 and 5 $R_{\odot}$ \citep{Hayes2001,Patoul_2015}. Hence, while integrating Equation \ref{eq:free_free_optical_depth}, we ignore the contributions to $n_\mathrm{e}$ from beyond 5 $R_{\odot}$. For a LoS with a plane-of-sky distance of $\sim$3 $R_{\odot}$, this leads to a LoS depth of about 8 $R_{\odot}$ within a sphere of 5 $R_{\odot}$. Considering these average values, $\tau_\nu$ becomes unity at $\nu\approx$ 31 MHz. The MWA observing frequency ranges from 80 - 240 MHz, which is many times higher than this value. Hence, the medium is not optically thick and rules out the optically thick free-free emission as one of the possible emission mechanisms. 

The brightness temperature ($T_\mathrm{B}$) of optically thin free-free emission is proportional to $\nu^{-2}$. This implies a flat flux density spectrum. $T_\mathrm{B}$ for optically thin free-free emission can be written as \citep{Gopalswamy1992},
\begin{equation}
    T_\mathrm{B}=\frac{<n_\mathrm{e}>^2L}{5\ T_e^{0.5}\ \nu^2}
    \label{eq:freefree_Tb}
\end{equation}
where, $<n_\mathrm{e}>$ is the average $n_\mathrm{e}$ along the LoS and $L$ is the LoS depth. Using this expression, we have estimated optically thin $T_\mathrm{B}$ is $\sim1235$ K. The rms $T_\mathrm{B}$ of the image at 100 MHz is $\sim$1100 K. Hence, the contribution from optically thin free-free emission from coronal plasma is below our detection limit.

The only likely emission mechanism remaining is the GS emission. In this instance, the faint GS emission from the CME plasma is detected out to a heliocentric distance of 8.3 $R_\odot$, the largest distance to which detection of such emission has been reported to date.

As is evident from Figure \ref{fig:south_cme}, the radio emission is not coming from the leading edge of the CME-2 (marked by the yellow curved line), but from behind it. Based on an examination of LASCO-C2 running difference images it has been found that as the CME passes, the pre-existing streamer structures are disturbed, but they are not completely disrupted even after the CME has moved out. Disturbances in magnetic fields are seen in the LASCO C2 base difference movie as the CME-2 passes over the region marked by red boxes in the static Figure \ref{movie:CME2}. This static figure shows some snapshots of the movie with increasing time from the top-left to the bottom-right corner. The first row shows three snapshots before CME-2 appeared in the LASCO C2 FoV. In these three frames, the streamer marked by black arrows maintains its morphology and shows two distinct sharp features. The dividing region of these two distinct features is marked by a blue dotted line in the first panel. In the second row, three snapshots are shown while CME-2 is passing the streamer region. The faint leading edge of the CME is marked by yellow dashed lines. During this time the streamer starts changing its morphology, becoming wider and more diffused. The third row shows snapshots after CME-2 has passed the streamer region. In these three snapshots, the two distinct components of the streamer almost merge and make a much wider diffused structure. These signatures show disturbances in the magnetic field structures of the streamer region after the CME-2 has passed over there. 

Disturbed magnetic field structures are visible inside the red box as CME-2 passes that region. Similar density enhancements and disturbances in streamer structures are also seen in STEREO-A COR2 base difference images.

We conjecture that the disturbed magnetic field from this region of interaction between the CME and the pre-existing streamer leads to some magnetic reconnection activity, which, in turn, produces mildly relativistic electrons. These electrons in the presence of the magnetic fields of that region give rise to the observed GS radio emission. With the currently available observations, there is no other independent way to confirm the presence of nonthermal electrons at these high coronal heights.

\subsection{Robust Detection of Circularly Polarized Radio Emission Associated With CME-2}\label{subsec:circular_pol}
Unlike most of the previous studies \citep{bastian2001,Carley2017,Mondal2020a}, this study presents high-fidelity full Stokes imaging of GS emission from a CME. The quality of polarimetric calibration and imaging provided by P-AIRCARS is comparable to high-quality astronomical observations \citep{Kansabanik2022_paircarsI}, which typically provides residual Stokes I to V leakage of $\sim0.1\%$. 

\begin{figure*}[!ht]
    \centering
    \includegraphics[trim={2.3cm 0cm 1cm 0cm},clip,scale=0.74]{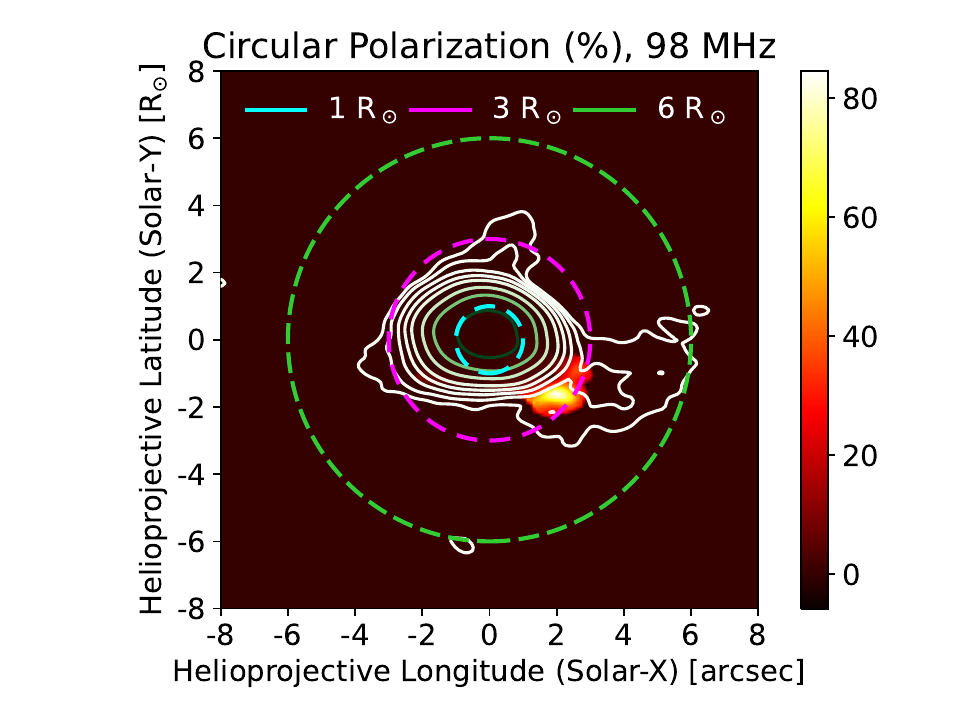}\includegraphics[trim={2.2cm 0cm 1cm 0cm},clip,scale=0.74]{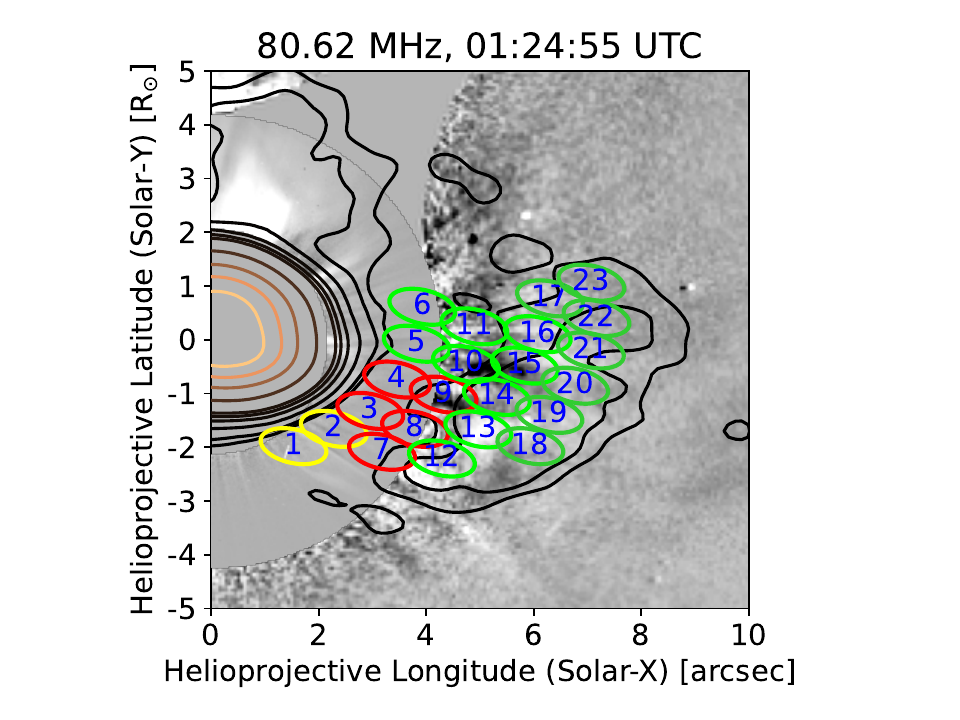}
    \caption{\textbf{Left panel:} Circular polarization image at 98 MHz. Background colormap shows percentage circular polarization and the contours represent the Stokes I emission. Contours at 0.5, 1, 2, 4, 8, 20, 40, 80 \% level of the peak Stokes I flux density. Three dashed circles are at 1, 3, and 6 $R_\odot$ and are shown by cyan, magenta, and green colors, respectively. \textbf{Right panel:} Regions of northern CME where spectra have been extracted. Contours at 0.5, 0.8, 2, 4, 6, 8, 20, 40, 60, and 80 \% levels of the peak Stokes I at 80 MHz are overlaid on the LASCO C2 and C3 coronagraph images. Red regions are those where spectrum fitting is done. Spectrum modeling is not done for green regions. Spectrum fitting is also done for yellow regions, which also have Stokes V detection at 98 MHz.}
    \label{fig:circular_pol_regions}
\end{figure*}

K23 presented a study of CME-1 which did not have a Stokes V detection from the CME plasma, but the authors placed stringent upper limits on it. Unlike K23, Stokes V emission is detected with high significance over a small part of Stokes I emission associated with CME-2. Percentage Stokes V image at 98 MHz is shown as the left panel of the background color map in Figure \ref{fig:circular_pol_regions} and corresponding Stokes I emission is shown by contours. Stokes V emission rises above the detection threshold only at 98 MHz and only for the two PSF-sized regions marked by yellow ellipses in the right panel of Figure \ref{fig:circular_pol_regions}. For other frequencies and other regions, we only have the stringent Stokes V upper limits estimated following the method used in K23 and briefly discussed later in Section \ref{sec:spectra_model}. Residual instrumental leakage for Stokes V is $<|1\%|$. The average polarization fraction detected over the regions marked by yellow ellipses in the right panel of Figure \ref{fig:circular_pol_regions} is $\sim50\%$. It is more than an order of magnitude larger than the residual instrumental polarization leakage and clearly establishes the robustness of the Stokes V detection.

\subsection{Spatially Resolved Spectroscopy}\label{subsec:spectroscopy_cme2}
We performed spatially resolved spectroscopy of the radio emission from CME-2 using wideband spectropolarimetric imaging observations with the MWA. Images are convolved with the PSF of the lowest observing frequency, 80 MHz. Then, spectra are extracted from the same PSF regions. These PSF-sized regions are marked in the right panel of Figure \ref{fig:circular_pol_regions}. For flux measurements at a given frequency to be considered reliable, we have followed the same criteria as used by K23. All of the criteria must be satisfied to consider a spectral point as detection. These criteria are as follows:
\begin{enumerate}
    \itemsep 0em 
    \item $f>\mu+5\sigma$
    \item $f>5\alpha$
    \item $f>5|n|$,
\end{enumerate}
where $f$ is the flux density of a PSF-sized region, $n$ is the deepest negative close to the Sun, $\sigma$ and $\mu$ are the rms noise and mean, respectively, calculated over a region close to the Sun, and $\alpha$ is the rms noise estimated far away from the Sun. These stringent criteria ensure that any contamination due to possible imaging artifacts is small. Uncertainty on Stokes I flux density is denoted by $\sigma_I$ and on Stokes V by $\sigma_V$. Spectra are fitted for the regions marked in red and yellow in the right panel of Figure \ref{fig:circular_pol_regions} which meet the detection criteria in more than five spectral bands. Although radio emission is detected up to 240 MHz, at high-frequency bands the emissions cover a region smaller than the size of the PSF at 80 MHz. For this reason, these spectral points are not included in further analysis even though they satisfy the above three criteria. 

\section{Spectral Modeling Using Homogeneous Source Model}\label{sec:spectra_model}
To date, all modeling of the GS radio emissions from CMEs has been done assuming a homogeneous source model along the LoS \citep[e.g.,][etc.]{bastian2001,Bain2014,Tun2013,Mondal2020a,Kansabanik2023_CME1}. Such a homogeneous GS source is populated with mildly relativistic electrons following a single power-law energy distribution. This simple GS model already has ten independent parameters -- magnetic field strength ($|B|$), angle between the LoS and the magnetic field ($\theta$), area of emission ($A$), LoS depth through the GS emitting medium ($L$), temperature ($T$), thermal electron density ($n_{thermal}$), non-thermal electron density ($n_{nonth}$), power-law index of non-thermal electron distribution ($\delta$),  $E_\mathrm{min}$, and $E_\mathrm{max}$. We have used the {\it ultimate fast GS code} by \cite{Kuznetsov_2021} for GS modeling.

K23 demonstrated that the model GS spectra are quite insensitive to variations in $T$ and $E_\mathrm{max}$. Hence, $T$ and $E_\mathrm{max}$ are kept fixed at 1 MK and 15 MeV respectively. $n_\mathrm{thermal}$ is estimated from inversion of the white light coronagraph images \citep{Hayes2001}. $|B|$, $\theta$, $A$, $\delta$ and $E_\mathrm{min}$ are fitted, while setting $n_\mathrm{nonth}$ to 1\% of the $n_\mathrm{thermal}$, similar to what has been assumed in earlier works \citep{Carley2017,Mondal2020a,Kansabanik2023_CME1}. $L$ is explicitly fitted only for region 3 which has seven spectral points available. Its upper limit is set to $L_\mathrm{max}$ obtained from 3D reconstruction from multi-vantage point white-light observations as described in Section \ref{subsec:geometrical_params}.

The mathematical framework used for joint spectral fitting using Stokes I detection and Stokes V upper limits has been discussed in detail in K23 and is summarized very briefly here. 
Bayes theorem is used \citep{Puga2015,Andreon2015_bayes_thereom} to estimate the posterior distribution, $\mathcal{P(\lambda|\mathcal{D})}$ of model parameters, $\lambda$, given the data, $\mathcal{D}$, and a likelihood function, $\mathcal{L(\mathcal{D}|\lambda)}$. When either Stokes I or Stokes V detections are used, the likelihood function is defined as
\begin{equation}
\begin{split}
      \mathcal{L}_\mathrm{1}(\mathcal{D}|\lambda)&=\mathrm{exp}\left(-\frac{1}{2}\sum_{i=1}^N \left[\frac{\mathcal{D}_\mathrm{i}-m_\mathrm{i}(\lambda)}{\sigma_\mathrm{i}}\right]^2\right)\\
      &=\prod_{i=1}^N \mathrm{exp}\left(-\frac{1}{2}\left[\frac{\mathcal{D}_\mathrm{i}-m_\mathrm{i}(\lambda)}{\sigma_\mathrm{i}}\right]^2\right),
\end{split}
\label{eq:likelihood_1_cme2}
\end{equation}
where $N$ is the total number of data points, $\mathcal{D}_\mathrm{i}$, $m_\mathrm{i}(\lambda)$, and $\sigma_\mathrm{i}$ are the observed values, models values and uncertainty on the measurements respectively. For the case of upper limits, the likelihood function is defined as follows \citep{Ghara2020,Greig2021,Maity2022}
\begin{equation}
\begin{split}
      \mathcal{L}_\mathrm{2}(\mathcal{D}|\lambda)&=\prod_{i=1}^N \frac{1}{2}\left[1-erf\left(\frac{\mathcal{D}_\mathrm{i}-m_\mathrm{i}(\lambda)}{\sqrt{2}\sigma_\mathrm{i}}\right)\right],
\end{split}
\label{eq:likelihood_2_cme2}
\end{equation}
where $erf$ refers to the error function. When both detections and upper limits are available, one can define the joint likelihood function as,
\begin{equation}
    \mathcal{L}(\mathcal{D}|\lambda)=\mathcal{L}_\mathrm{1}(\mathcal{D}|\lambda)\ \mathcal{L}_\mathrm{2}(\mathcal{D}|\lambda),
    \label{eq:join_likelihood_cme2}
\end{equation}
which allows one to use the constraints from the detections as well as the upper limits. We use the Monte Carlo Markov Chain \citep[MCMC,][]{brooks2011handbook} analysis to estimate the posterior distribution of parameters using the joint likelihood function. Although Stokes V detection was reported earlier \citep{bastian2001,Tun2013}, for the first time, we have jointly used Stokes V detection with Stokes I spectra to constrain the GS model.

\subsection{Estimation of Geometrical Parameters}\label{subsec:geometrical_params}
\begin{figure*}[!ht]
\centering 
    \includegraphics[trim={0.5cm 4.5cm 0cm 5cm},clip,scale=0.45]{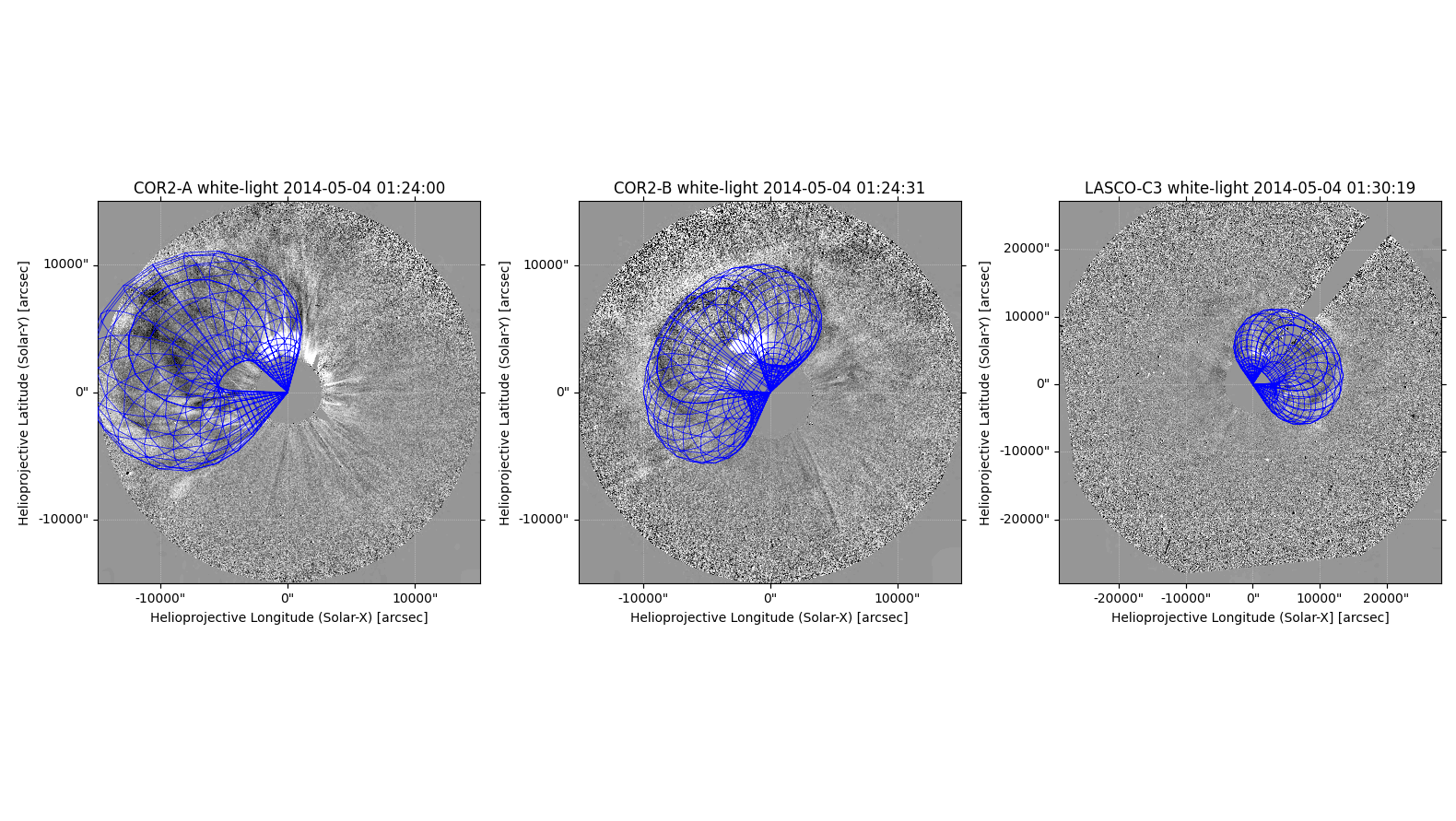}\\   
    \includegraphics[trim={0cm 0cm 0cm 0cm},clip,scale=0.15]{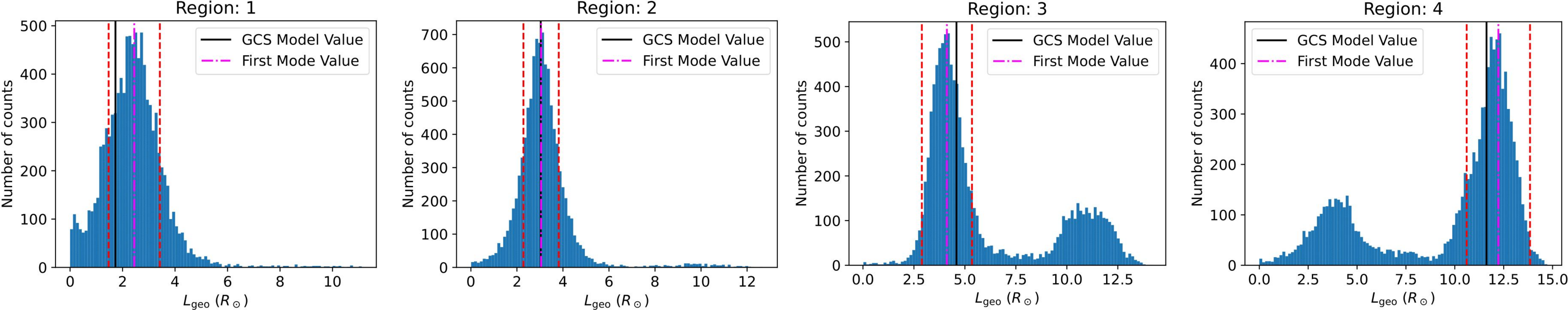}
    \caption{\textbf{Top row:} Three-dimensional reconstruction of the CME-2 using Graduated Cylindrical Shell (GCS) model. GCS fitting is constrained using three vantage point observations from the LASCO-C2, C3, and COR-2 coronagraphs onboard STEREO-A and STEREO-B spacecraft. Blue meshes show different views of the GCS model of CME-2 at 01:30 UTC on 2014 May 04. \textbf{Bottom row:} Distributions of $L_\mathrm{geo}$ for some sample PSF-sized regions of CME-2. Solid black lines represent $L_\mathrm{geo}^{GCS} (R_\odot)$} for the GCS model parameters listed in Section \ref{subsec:geometrical_params}. Dash-dot magenta lines represent the mode and red dashed lines represent the median absolute deviation around the mode. The mode and corresponding standard deviations are mentioned in Table \ref{table:los_depth_cme2}.
    \label{fig:gcs}
\end{figure*}

\begin{table*}[!ht]
\centering
    \renewcommand{\arraystretch}{1.4}
    \begin{tabular}{|p{1.4cm}|p{1.8cm}|p{1.8cm}|p{2.2cm}|p{1.4cm}|p{1.8cm}|p{1.8cm}|p{2.2cm}|}
    \hline
       Region & $L_\mathrm{geo} (R_\odot)$ & $L_\mathrm{geo}^{GCS} (R_\odot)$ & $s\sigma(L_\mathrm{geo})\ (R_\odot)$ & Region & $L_\mathrm{geo}\ (R_\odot)$  & $L_\mathrm{geo}^{GCS} (R_\odot)$ & $\sigma(L_\mathrm{geo})\ (R_\odot)$\\ \hline \hline 
        1 & 2.45 & 1.73 & 0.98 &  7 & 3.57 & 4.0 & 1.25\\
        \hline
        2 & 3.05 & 3.04 & 0.77 &  8 & 4.68 & 5.13 & 3.52\\
        \hline
        3 & 4.12 & 4.59 & 1.23 &  9 & 11.94 & 11.82 & 1.27\\
        \hline
        4 & 12.23 & 11.62 & 1.62 & -- & -- & -- & --\\
       \hline
    \end{tabular}
    \caption{\textbf{Estimated geometric LoS depth from GCS modeling of south-western CME.} The geometric LoS depths are obtained for red regions using ray tracing from Earth through that region. Geometric LoS depths are given in units of the solar radius.}
    \label{table:los_depth_cme2}
\end{table*}
\begin{table*}[!htpb]
\centering
    \renewcommand{\arraystretch}{1.5}
    \begin{tabular}{|p{1cm}|p{1.6cm}|p{1.3cm}|p{1.5cm}|p{1.5cm}|p{1.8cm}|p{1.8cm}|p{1.5cm}|p{1.2cm}|p{1.2cm}|}
    \hline
       Region No. & Heliocentric \newline{Distance} & $|B|$ (G) & $\delta$ & $ A \times 10^{20}$\newline{$(cm^{2})$} & $E_\mathrm{min}$ (keV) & $\theta$ \newline{(degrees)} & $L\ (R_\odot)$ & $n_\mathrm{thermal}$ \newline{$\times 10^6$}\newline{$(cm^{-3})^*$}  & $n_\mathrm{nonth}$\newline{$\times 10^4$} \newline{$(cm^{-3})^*$} \\ \hline \hline 
        1 & 2.65 & $0.20_{-0.02}^{+0.02}$ & $2.59_{-0.37}^{+0.53}$ & $2.06_{-0.93}^{+1.08}$ & $278.74_{-148.64}^{+293.77}$ & $38.41_{-2.86}^{+4.31}$  & $1.42^*$ &1.25 & 1.25\\
        \hline
        2 & 2.65 & $1.72_{-0.52}^{+0.64}$  & $6.82_{-1.56}^{+1.88}$  & $4.76_{-1.84}^{+2.94}$  & $139.65_{-67.09}^{+107.86}$  & $79.34_{-10.22}^{+5.96}$ & $1.41^*$ & 1.25 & 1.25\\
        \hline
        3 & 2.65 & $3.99_{-0.76}^{+0.71}$ & $5.94_{-1.10}^{+1.89}$ & $10.24_{-3.39}^{+7.72}$ & $35.69_{-16.70}^{+26.07}$ & $77.67_{-5.69}^{+4.58}$ & $1.98_{-1.15}^{+2.04}$ & 1.25 & 1.25\\
        \hline
        4 & 2.65 & $1.44_{-0.44}^{+0.64}$ & $4.02_{-1.05}^{+2.34}$ & $0.94_{-0.47}^{+1.15}$ & $67.77_{-40.94}^{+133.62}$ & $53.01_{-18.38}^{+23.68}$ & $5.12^*$ &1.25 & 1.25\\
       \hline
       7 & 3.2 & $1.61_{-0.86}^{+0.51}$ & $6.42_{-2.34}^{+2.52}$ & $2.88_{-2.17}^{+4.81}$ & $139^*$ & $71.12_{-17.83}^{+13.33}$ & $1.78^*$ & 0.7 & 0.7\\
       \hline
        8 & 3.2 & $2.36_{-0.56}^{+0.39}$ & $8.57_{-2.42}^{+2.81}$ & $10.27_{-6.93}^{+9.67}$ & $139^*$ & $64.73_{-12.45}^{+15.70}$ & $3.03^*$ & 0.7 & 0.7\\
       \hline
        9 & 3.2 & $2.49_{-0.49}^{+0.66}$ & $6.51_{-0.86}^{+1.98}$ & $5.30_{-2.27}^{+8.68}$ & $67^*$ & $63.06_{-12.19}^{+17.10}$ & $4.88^*$ & 0.7 & 0.7\\
       \hline
    \end{tabular}
    \caption{\textbf{Estimated GS model parameters of CME-2 considering homogeneous GS source model.} These parameters are estimated for 01:24:55 UTC. Parameters marked by $^*$ are kept fixed during the fitting.}
    \label{table:south_params}
\end{table*}
Multiple vantage point observations using SOHO, STEREO-A, and STEREO-B spacecraft allow us to perform a 3D reconstruction of CME-2. 3D reconstruction is performed using the Graduated Cylindrical Shell \citep[GCS;][]{Thernisien_2006,Thernisien_2011} model using its {\it Python} implementation \citep{gcs_python}. We obtained a good visual fit to the GCS model using the white-light images following the method described by \cite{Thernisien2009}. The GCS model at 01:30 UTC is shown by blue mesh in the top row of Figure \ref{fig:gcs}, where different panels show superposition on COR-2 images from STEREO-A and STEREO-B, and C3 coronagraph images from LASCO. The best visual fit GCS model parameters at 01:30 UTC are:
\begin{enumerate}
    \itemsep 0em 
    \item Front height ($h_{front}$) : 17.9 $R_\odot$
    \item Half-angle ($\alpha$) : 46.8$^{\circ}$
    \item Carrington Longitude ($\Phi$) : 36.5$^{\circ}$ 
    \item Heliospheric Latitude ($\Theta$) : 14.1$^{\circ}$ 
    \item Aspect Ratio ($\kappa$) : 0.45
    \item Tilt Angle ($\gamma$) : -56.7$^{\circ}$
\end{enumerate}
\subsubsection{Estimating $L_\mathrm{geo}$ from GCS Model}\label{subsubsec:los_depth_estimate_south}
We performed ray-tracing through the GCS shell and computed the geometrical path length ($L_\mathrm{geo}$) intersected by the GCS shell of CME-2 for each PSF-sized region using {\it Python}-based ray-tracing code {\it trimesh} \citep{trimesh}. All rays originate from the Earth. The length of the ray segment inside the GCS shell is regarded as $L_\mathrm{geo}$ for that PSF-sized region. Since, $L_\mathrm{geo}$ is estimated using the numerical method of ray-tracing, there is no analytic relationship between GCS model parameters and $L_\mathrm{geo}$. Hence, it is not possible to estimate uncertainty on $L_\mathrm{geo}$ ($\sigma(L_\mathrm{geo})$) using usual error propagation. To estimate the uncertainty on $L_\mathrm{geo}$, we have used an approach of ensemble modeling. We have generated 10,000 realizations of GCS model parameters considering independent Gaussian distributions for each of these GCS parameters. The mean of these Gaussian distributions was set to the fitted values obtained from the visual fit. The standard deviation was set to the uncertainty on GCS model parameters reported in \cite{VERBEKE2022} based on analysis of several synthetic CMEs and multi-spacecraft observations. We have then computed the GCS shells for all combinations of these GCS parameters drawn from Gaussian distributions and $L_\mathrm{geo}$ was computed for each of these realizations. $L_\mathrm{geo}$ for the GCS parameters mentioned in Section \ref{subsec:geometrical_params} are denoted by $L_\mathrm{geo}^{GCS}$.

\begin{figure*}[!htbp]
    \centering
    \includegraphics[trim={0.3cm 0.5cm 0.0cm 0.3cm},clip,scale=0.37]{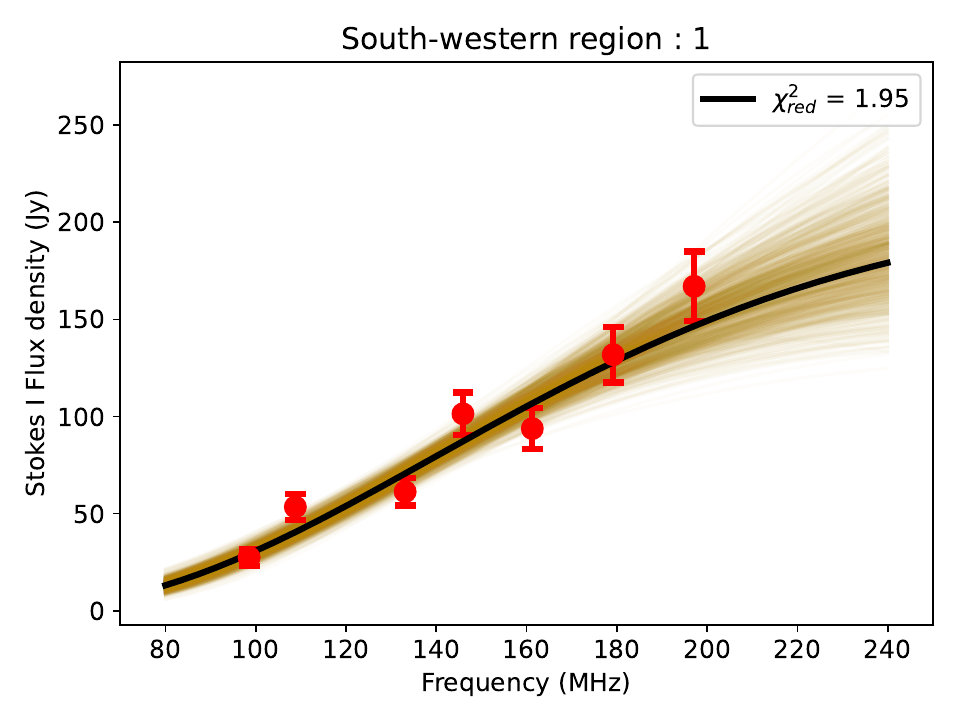}\includegraphics[trim={0.3cm 0.5cm 0.0cm 0.3cm},clip,scale=0.37]{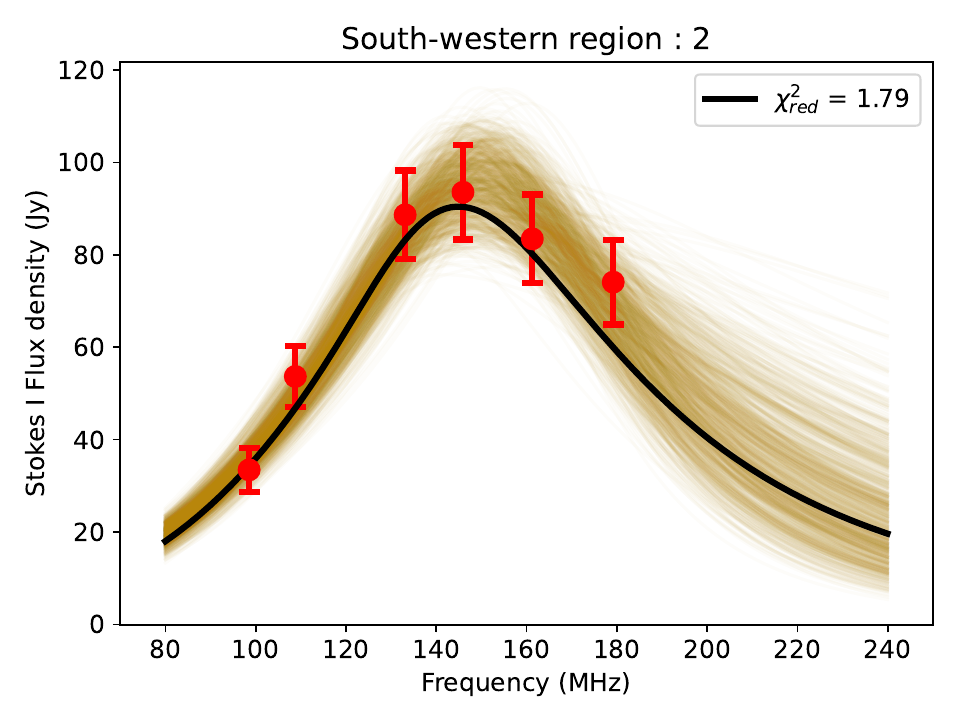}\includegraphics[trim={0.3cm 0.5cm 0.0cm 0.3cm},clip,scale=0.37]{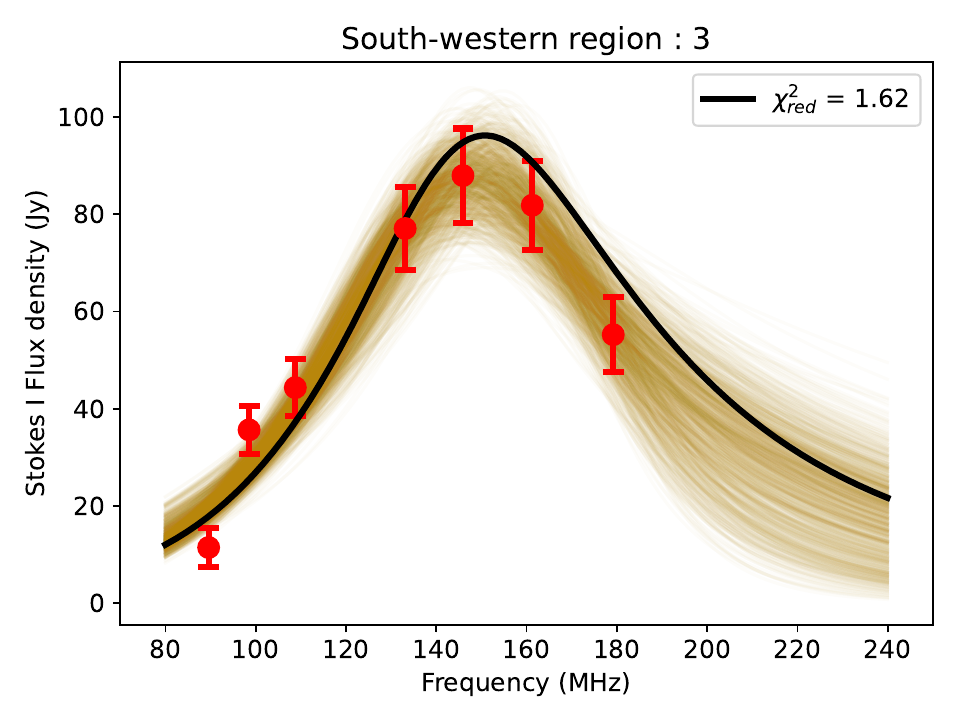}
 
     \includegraphics[trim={0.3cm 0.5cm 0.0cm 0.3cm},clip,scale=0.37]{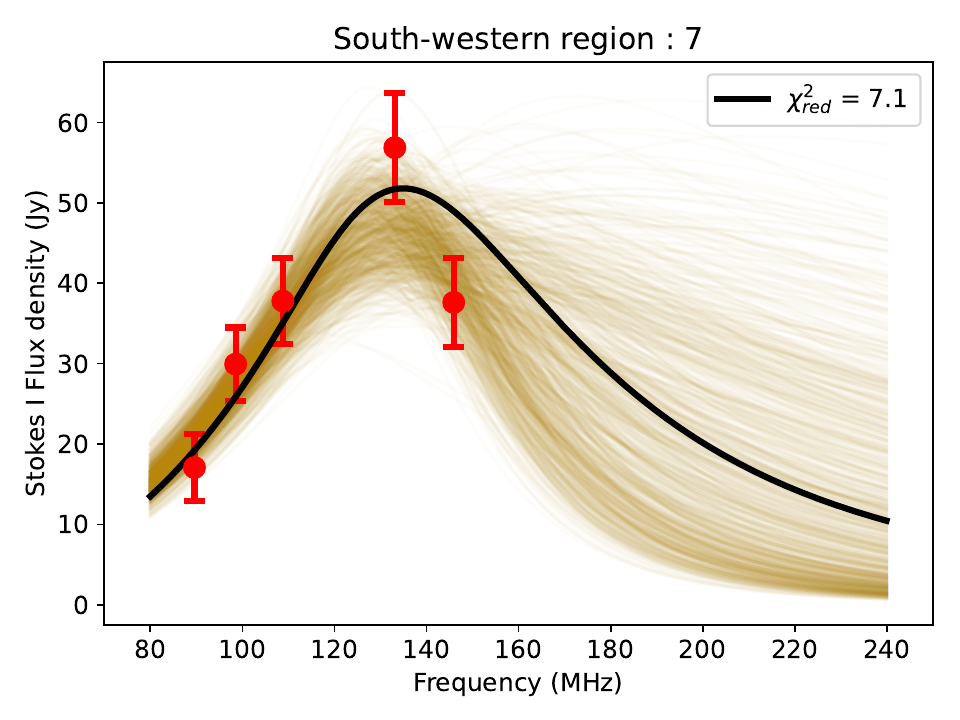}\includegraphics[trim={0.3cm 0.5cm 0.0cm 0.3cm},clip,scale=0.37]{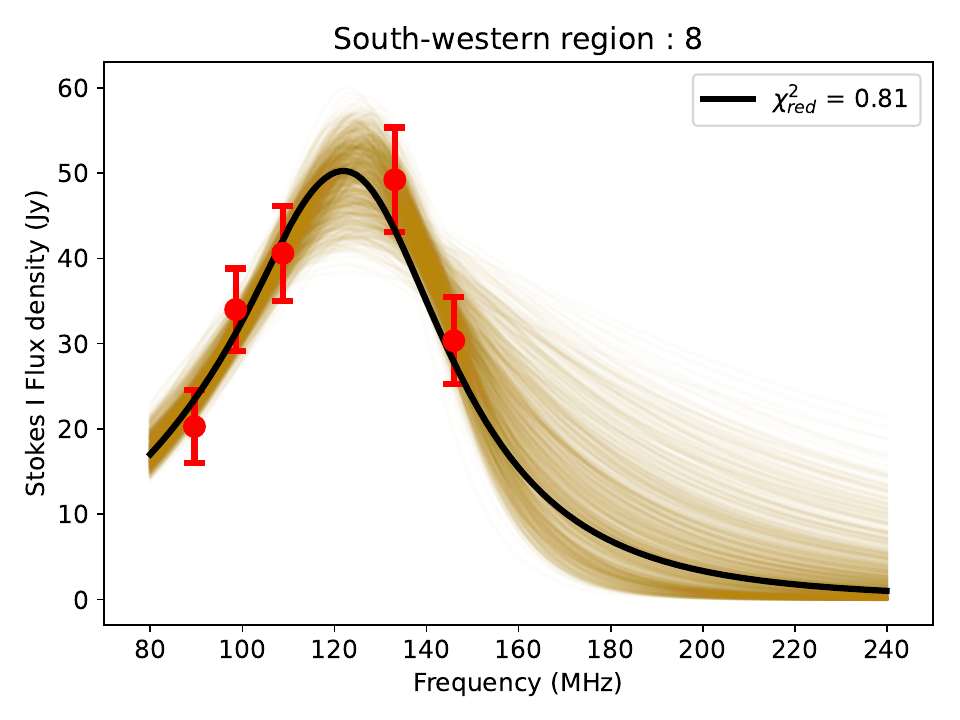}\includegraphics[trim={0.3cm 0.5cm 0.0cm 0.3cm},clip,scale=0.37]{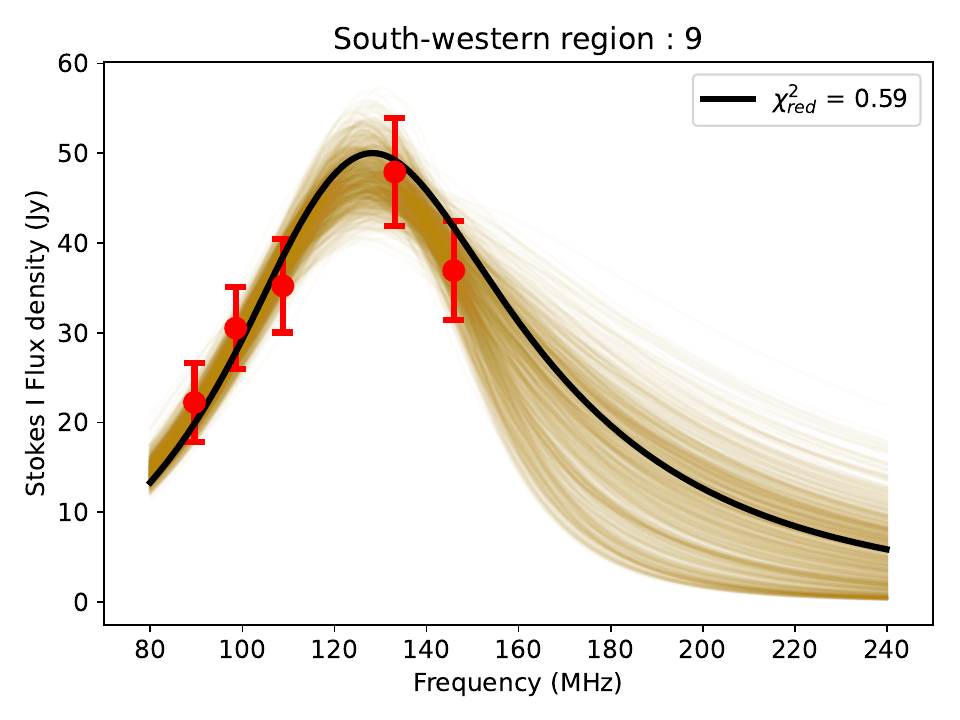}

      \includegraphics[trim={0.3cm 0.5cm 0.0cm 0.3cm},clip,scale=0.37]{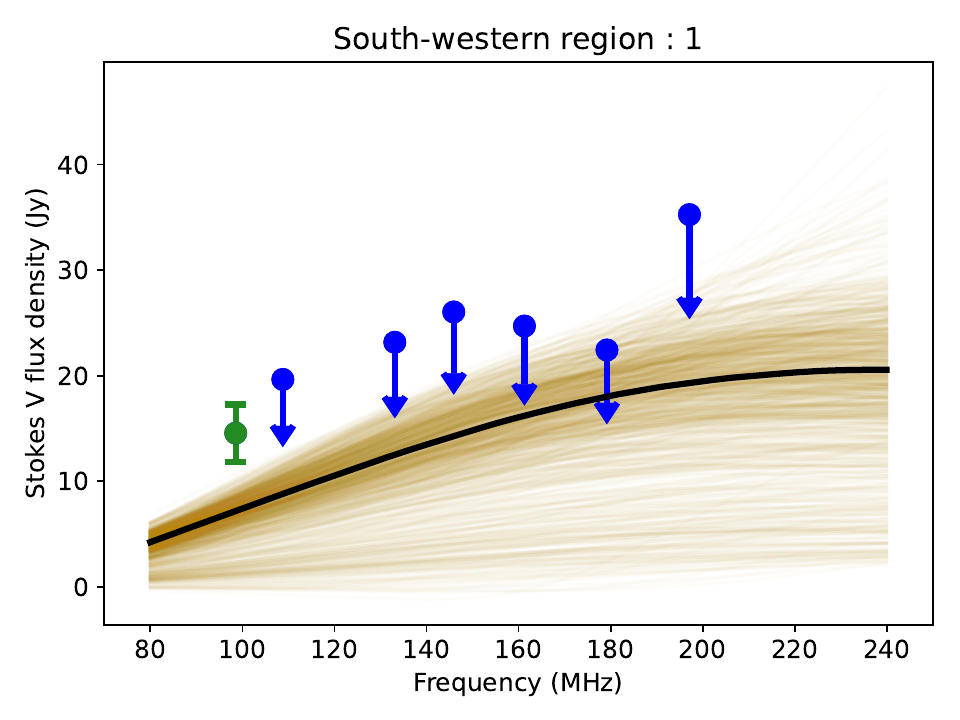}\includegraphics[trim={0.3cm 0.5cm 0.0cm 0.3cm},clip,scale=0.37]{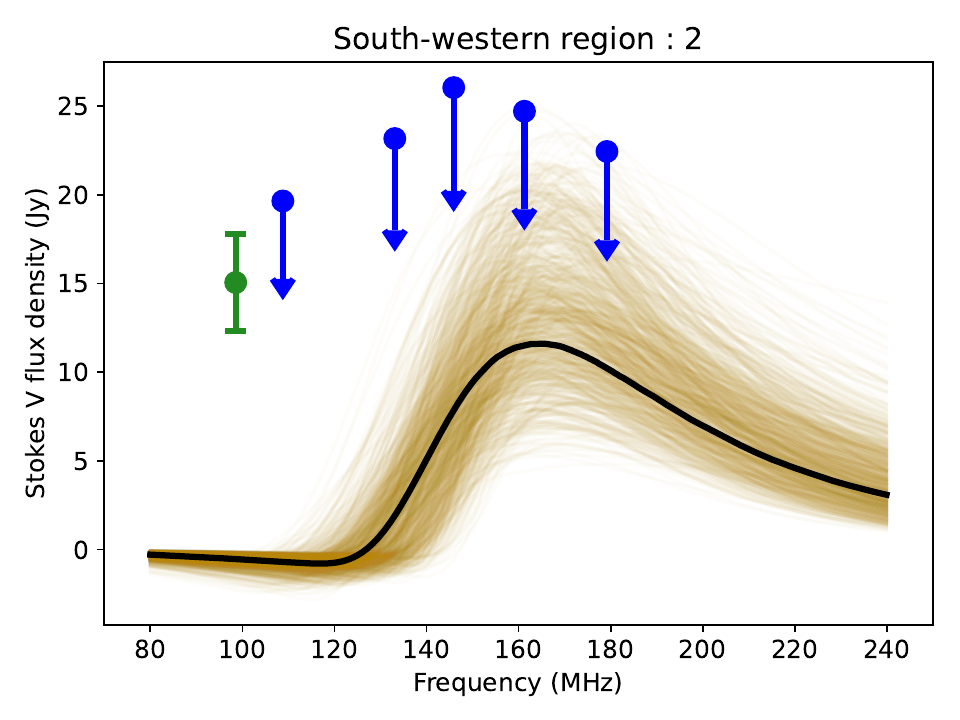}\includegraphics[trim={0.3cm 0.5cm 0.0cm 0.3cm},clip,scale=0.37]{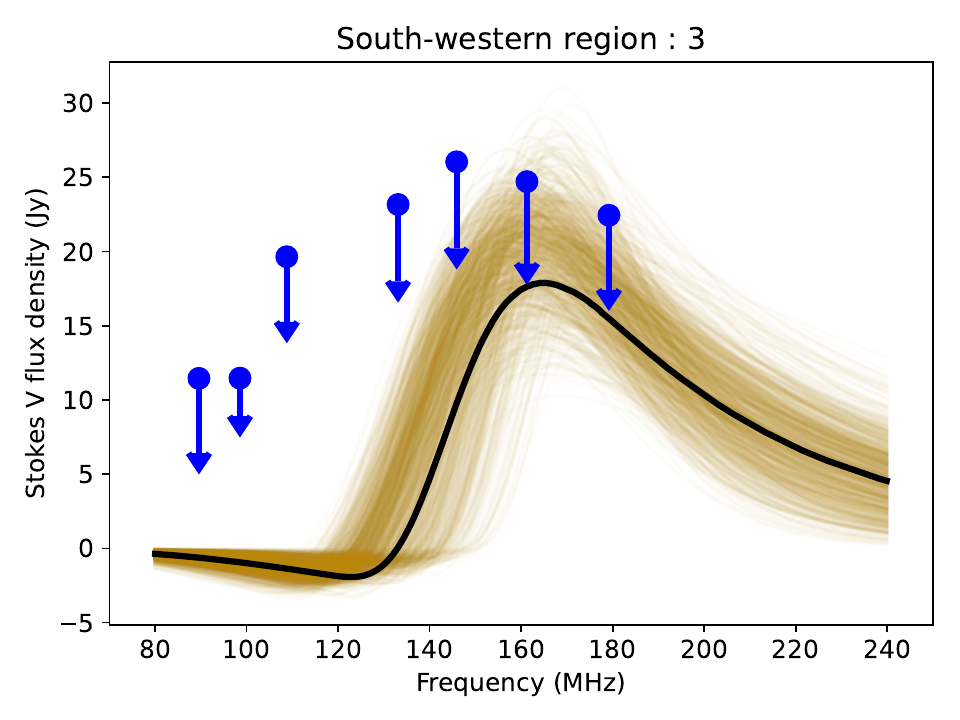}

     \includegraphics[trim={0.3cm 0.5cm 0.0cm 0.3cm},clip,scale=0.37]{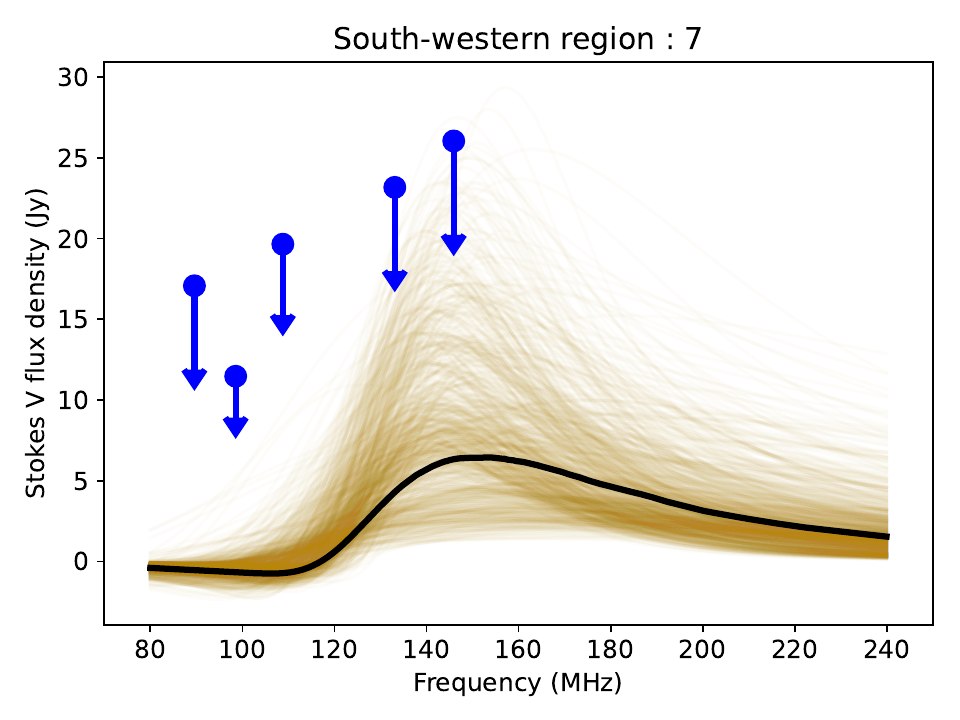}\includegraphics[trim={0.3cm 0.5cm 0.0cm 0.3cm},clip,scale=0.37]{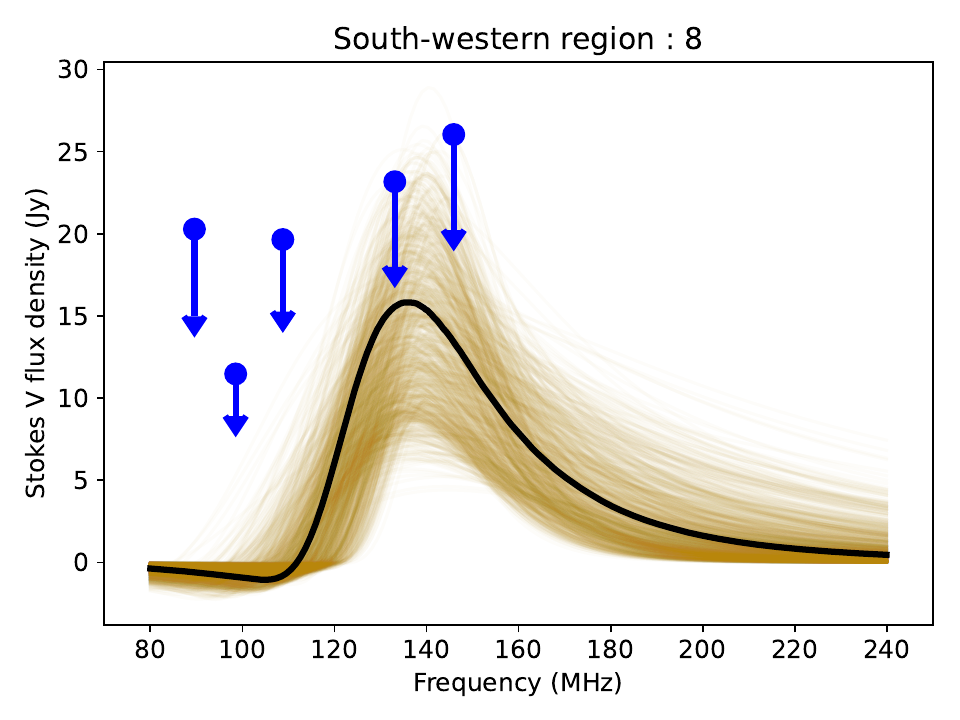}\includegraphics[trim={0.3cm 0.5cm 0.0cm 0.3cm},clip,scale=0.37]{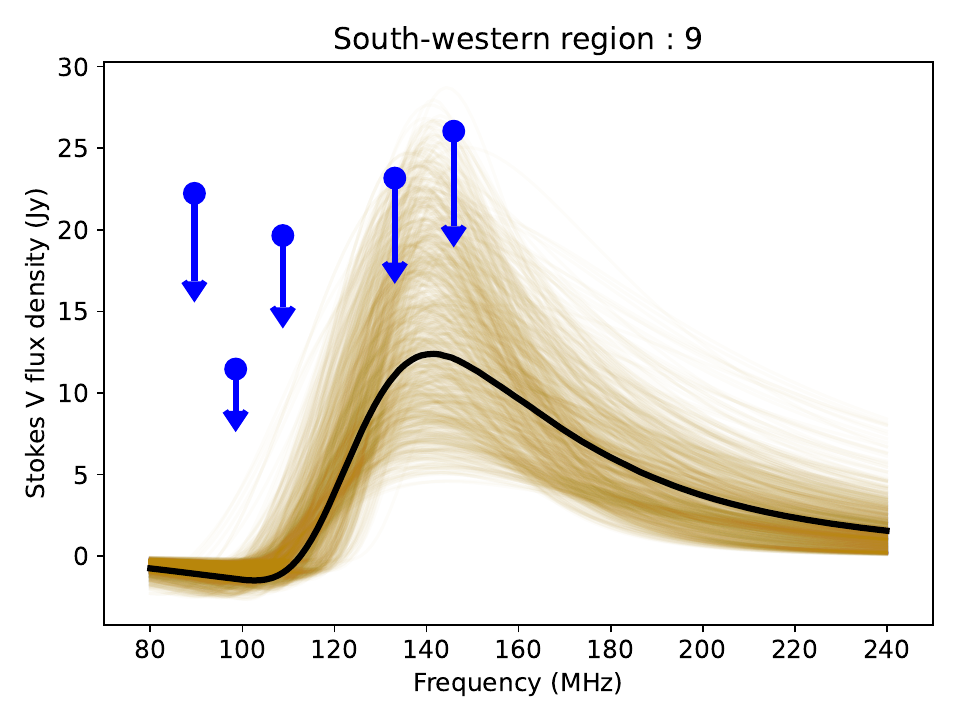}
    \caption{\textbf{Observed and fitted spectra for regions 1, 2, 3, 7, 8, and 9 of CME-2. First and second rows: }Stokes I spectra are shown. Red points represent the observed flux densities. \textbf{Third and fourth rows: }Stokes V spectra are shown. Blue points represent the observed upper limits. Green points show the Stokes V detections. The black lines represent the GS spectra corresponding to GS parameters reported in Table \ref{table:south_params}. Light yellow lines show the GS spectra for 1000 randomly chosen realizations from the posterior distributions of the GS model parameters resulting from a total of 1,000,000 MCMC chains.}
    \label{fig:spectra_homo1}
\end{figure*}
\begin{figure*}[!htbp]
    \centering
    \includegraphics[trim={0.6cm 0.6cm 0cm 0cm},clip,scale=0.6]{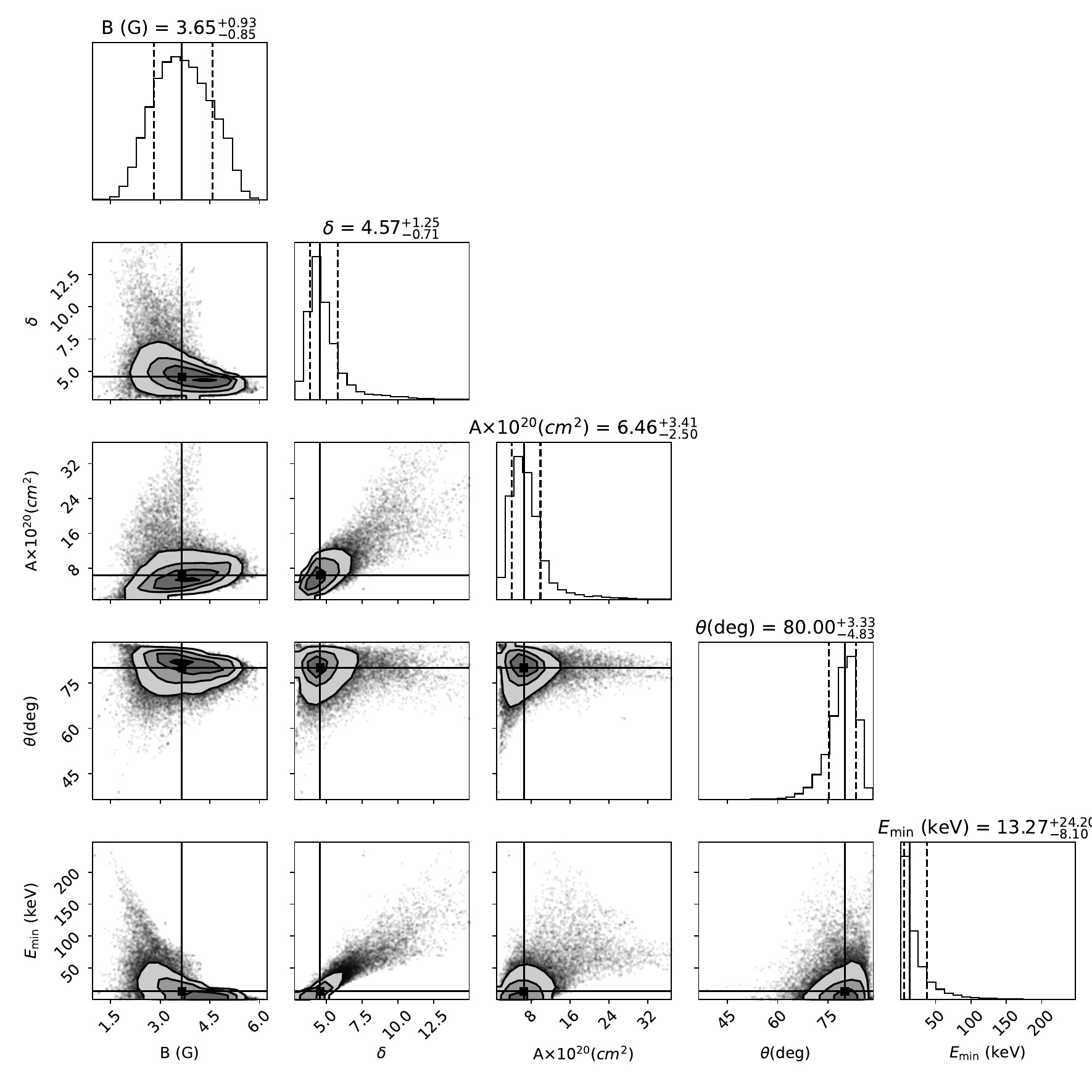}
    \caption{\textbf{Correlation of posterior distributions of GS model parameters for region 2.} 2-dimensional plots show the joint probability distribution of any two parameters. The contours are at 0.5, 1, 2, and 3$\sigma$. The solid lines in the 1-dimensional histogram of posterior distributions mark the median values, and the vertical dashed lines mark the 16$^\mathrm{th}$ and 84$^\mathrm{th}$ percentiles. The median values are also marked in the panels showing the joint probability distribution.}
    \label{fig:corner_reg2}
\end{figure*}

While the histograms of distributions of $L_\mathrm{geo}$ for most of the PSF-sized regions show an unimodal distribution, those for regions 3 and 4 show a bimodal distribution with a secondary peak of much lower amplitude, as shown in the bottom row of Figure \ref{fig:gcs}. We found that the $L_\mathrm{geo}$ (marked by solid black lines in the bottom row of Figure \ref{fig:gcs}) corresponding to the GCS model parameters mentioned above lie close to the mode of the distribution (marked by a dash-dot magenta line in the bottom row of Figure \ref{fig:gcs}). Hence, instead of using the mean or the median, the mode value is used as $L_\mathrm{geo}$ and standard deviation, $\sigma(L_\mathrm{geo})$, is estimated as $1.4826\times$ MAD, where MAD is the ``median absolute deviation" with respect to mode value (assuming the distribution is quasi-Gaussian around the mode). $L_\mathrm{geo}\pm\sigma(L_\mathrm{geo})$ are shown by dashed red lines in the bottom row of Figure \ref{fig:gcs}. It is important to note that LoS depth in the GS model ($L$) could be different from $L_\mathrm{geo}$. $L_\mathrm{geo}$ and $\sigma(L_\mathrm{geo})$ are tabulated in Table \ref{table:los_depth_cme2}. The maximum value of $L$ for a given region is chosen to be $L_\mathrm{max}=L_\mathrm{geo}+\sigma(L_\mathrm{geo})$.

\subsection{Joint Spectral Fitting of Stokes I and V Spectra Using Homogeneous GS Model}\label{subsec:homo_modeling}
A joint spectral fitting is done using Stokes I and V spectra for all regions marked in red or yellow in the right panel of Figure \ref{fig:circular_pol_regions}. Uniform priors, $\pi(\lambda)$, used for the GS model parameters are:
\begin{enumerate}
    \itemsep0em
    \item $B\ (\mathrm{G})$ : $(0,\ 10]$
    \item $\theta\ (\mathrm{degree})$ : $(0,\ 90)$
    \item $\delta$ : $(1,\ 10]$
    \item $A \times10^{20}\ (\mathrm{cm^{2}})$ : $[0.0001,\ 100]$
    \item $E_\mathrm{min}\ (\mathrm{keV})$ : $(0.1,\ 100]$
    \item $L\ (R_\odot)$ : $(0.01,L_\mathrm{max}]$.
\end{enumerate}

Modeled and observed Stokes I and V spectra are shown in Figure \ref{fig:spectra_homo1}. Modeled spectra are consistent with the observed Stokes I spectra and Stokes V upper limits for regions 1, 2, 3, 7, 8, and 9 as evident from Figure \ref{fig:spectra_homo1}. GS model parameters are well-constrained as evident from the posterior distribution of GS model parameters for region 2 shown in Figure \ref{fig:corner_reg2}. 

The spectrum for Region 3 has seven Stokes I spectral points. Hence, for region 3, $L$ is kept as a free parameter. Fractional GS source depth, $f=L/L_\mathrm{max}$, for this region is $\sim0.37$. For other regions, there are less than seven spectral points. Hence, to keep the number of free parameters in check, $f$ for other regions is assumed to be similar to that for region 3, and $L$ is kept fixed at $L=f\times L_\mathrm{max}$ and mentioned in Table \ref{table:south_params}. For regions 7, 8, and 9, there are 5 spectral points. For these regions, $E_\mathrm{min}$ is also kept fixed at the values estimated for nearby regions. GS model parameters for all these regions are presented in Table \ref{table:south_params}. 

The importance of sampling the spectral peak of GS spectra for constraining the GS model parameters was demonstrated by K23. For regions 1 (shown in the top panel of Figure \ref{fig:spectra_homo1}) and 4 of CME-2, only the optically thick part of the spectra have been sampled. For these regions, GS model parameters are not well-constrained.

\section{Examining the Validity of Homogeneous and Isotropic Assumptions of GS Model}\label{subsec:homo_insuff}
Since the first attempt to model GS emission from CME loops to estimate plasma parameters and magnetic field by \cite{bastian2001}, all studies have assumed a homogeneous and isotropic GS source model. At the same time, no observational evidence has been reported to suggest that this assumption is not valid. The modeling of the observed spectra of CME-2 from regions 3, 7, 8, and 9 and CME-1 (presented in K23) including Stokes I spectra and stringent Stokes V upper limits also use homogeneous and isotropic GS models and yield well-constrained model parameters consistent with the observations. 

This is, however, not the case for regions 1 and 2 of CME-2. These two regions are different from others in that Stokes V emission has been detected from these regions at 98 MHz. Observed and modeled Stokes I and Stokes V spectra for regions 1 and 2 are shown in the first and third rows of Figure \ref{fig:spectra_homo1}, respectively. While the peak of the spectrum lies beyond the MWA frequency range for region 1, the spectral peak has been sampled well for region 2. More importantly perhaps, while it is possible to find GS models consistent with the Stokes I measurements and the Stokes V upper limits, there is no GS model in the entire phase space explored which is simultaneously also consistent with the lone Stokes V measurement at 98 MHz. The ranges of the physical parameters explored here are sufficiently wide and it would be hard to justify expanding them beyond their present values. {\it A situation like the present, where a good model fit can be found for less constraining data, but as the constraints become tighter, it is no longer possible to find a good model fit, strongly suggests the need to critically examine the possibility of one or more of the assumptions made by the model being violated.} To examine this possibility and attempt to identify the specific assumption being violated, we systematically examine these assumptions, one at a time in the remainder of this section. The key assumptions examined are -- restricting the electron energy distribution to a single power law, ignoring any anisotropy in the nonthermal electron pitch-angle distribution, and the assumption of homogeneity in the plasma present in the volume being modeled by the GS model.

\subsection{Possible Effects of Different Electron Energy Distributions}\label{subsec:assumptions1}
The {\it ultimate fast GS code} developed by \cite{Kuznetsov_2021} allows the flexibility to model GS emission using different analytical electron energy distribution functions -- single (PLW) and double power-law (DPL) distributions, thermal/nonthermal distribution over energy (TNT), isotropic thermal and power-law over energy (TPL) and isotropic thermal and double power-law over energy (TPD) \citep{Fleishman_2010}. We have considered these different electron energy distributions. The observed Stokes I and V spectra for region 2 are fitted jointly considering homogeneous and isotropic GS model with each of these electron distributions. DPL and TPD models require a larger number of free parameters. To keep the problem well constrained, the geometric parameters $A$ and $L$ are kept fixed at the value mentioned in Table \ref{table:south_params} for all three models considered. 

For all of these distributions, the modeled spectra are consistent with the Stokes I flux densities and Stokes V upper limits. However, the lone observed Stokes V detection is not consistent with any of the models. This exercise establishes that none of the prevalent homogeneous and isotropic electron density models can reproduce the observed Stokes I and V spectra simultaneously.

\subsection{Possible Effects of Anisotropic Electron Distribution}\label{subsec:anisotropy_effect}
The mildly-relativistic electrons injected in CME plasma either due to magnetic reconnection or shock acceleration could well be anisotropic \citep{DuBois2017,agudelo_2021} during the initial phases and the anisotropy can sustain till later times \citep{Simnett_2002,Giacalone_2021}. This anisotropic distribution becomes isotropic over time due to collisional or turbulent scattering \citep{Kuznetsov_2021}. Hence, it is interesting to consider the impact of an anisotropic pitch-angle distribution for modeling the observed GS spectra for region 2 of CME-2. We have considered two types of analytical pitch-angle distribution available in {\it ultimate fast GS code} \citep{Kuznetsov_2021}. These are -- Gaussian beam distribution (GAU) and Gaussian loss-cone distribution (GLC)\citep{Fleishman_2010}.

GS spectrum modeling is performed for region 2 of CME-2 considering homogeneous PLW electron distribution with both GLC and GAU pitch-angle distributions. The modeled spectra for GAU and GLC pitch-angle distributions are unable to reproduce the observed Stokes I and V simultaneously. 

\begin{figure*}[!htbp]
    \centering
    \includegraphics[trim={0cm 0cm 0cm 0cm},clip,scale=0.35]{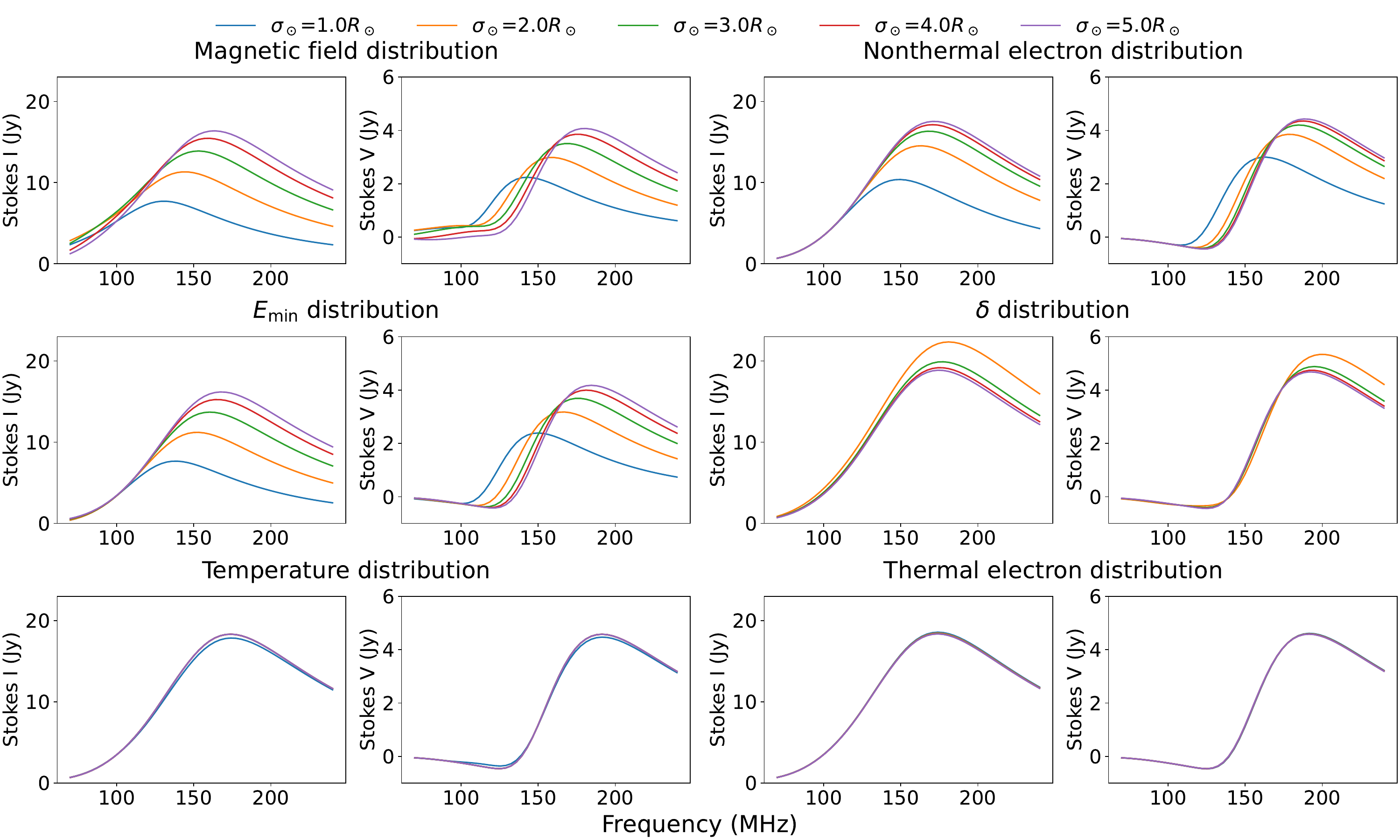}
    \caption{\textbf{Effects of inhomogeneous distributions of GS model parameters on simulated spectra. First and third columns: }Simulated Stokes I spectra. \textbf{Second and fourth columns:} Simulated Stokes V spectra. Different colors represent Gaussian distribution with different widths mentioned at the top of the figure.}
    \label{fig:simulation_1}
\end{figure*}
\begin{figure*}[!htbp]
    \centering
    \includegraphics[trim={0cm 0cm 0cm 0.5cm},clip,scale=0.45]{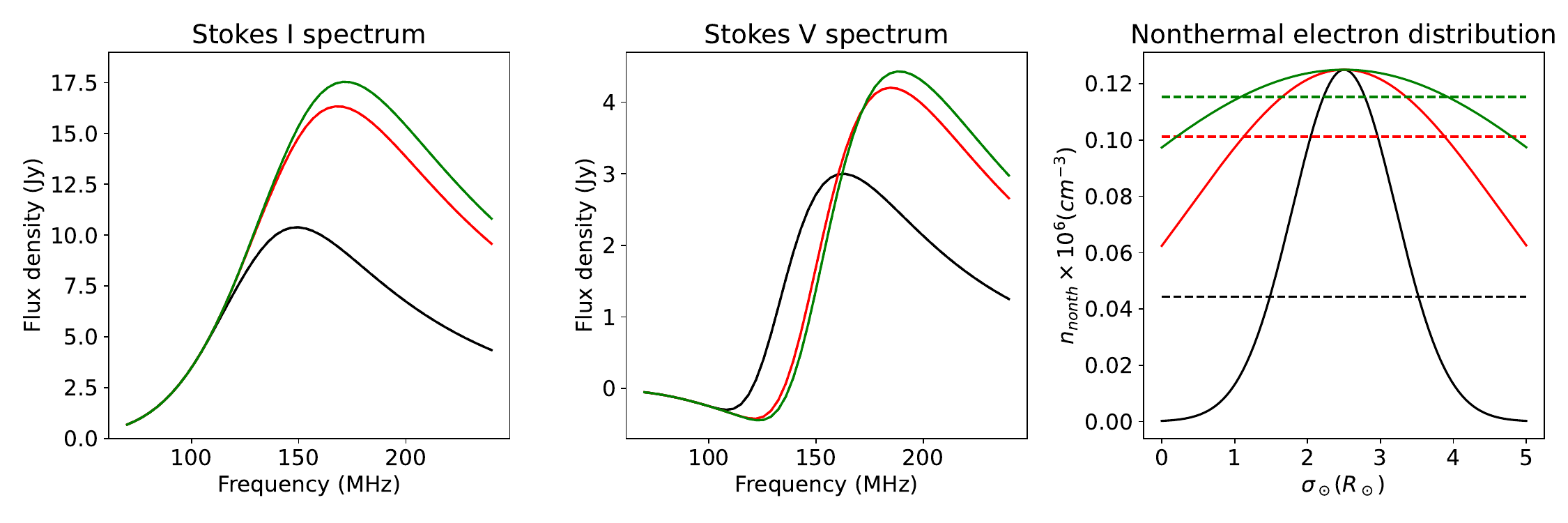}\\
    \includegraphics[trim={0cm 0cm 0cm 0.5cm},clip,scale=0.45]{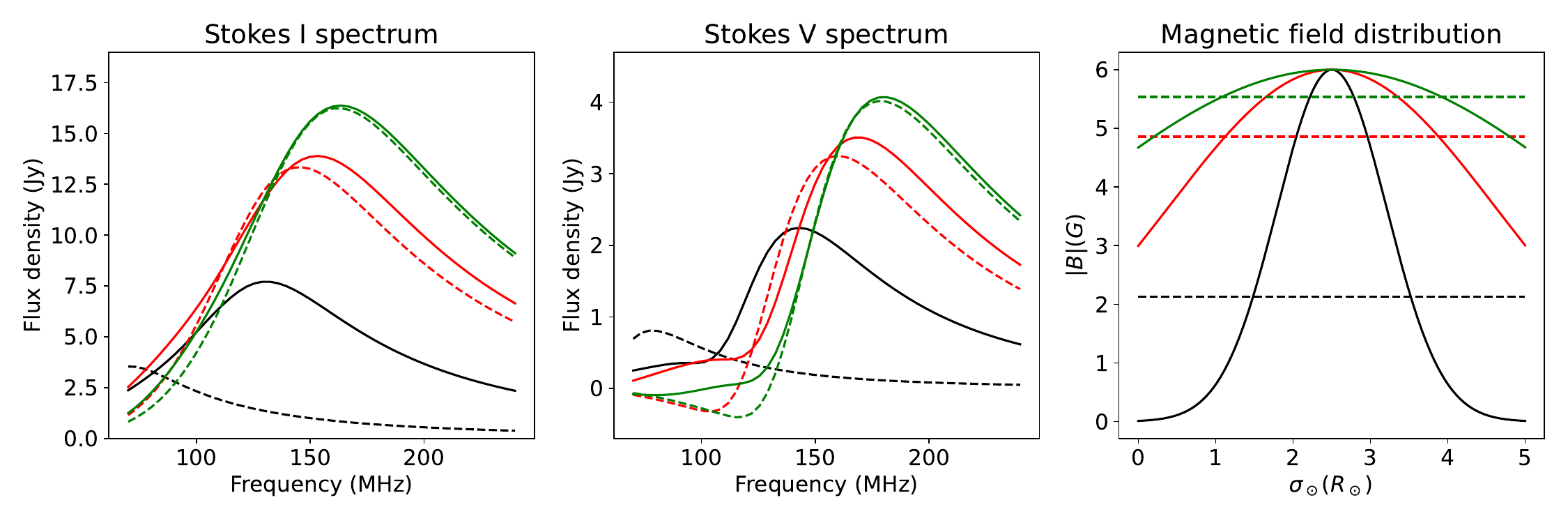}\\
    \includegraphics[trim={0cm 0cm 0cm 0.5cm},clip,scale=0.45]{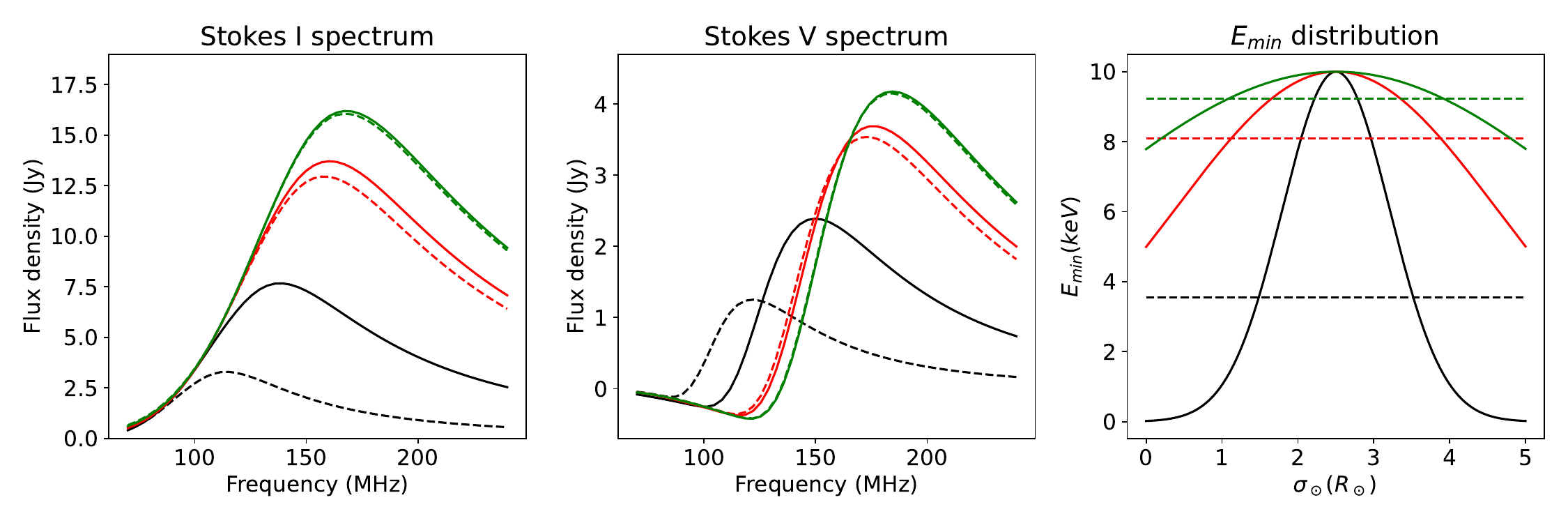}\\
    \includegraphics[trim={0cm 0cm 0cm 0.5cm},clip,scale=0.45]{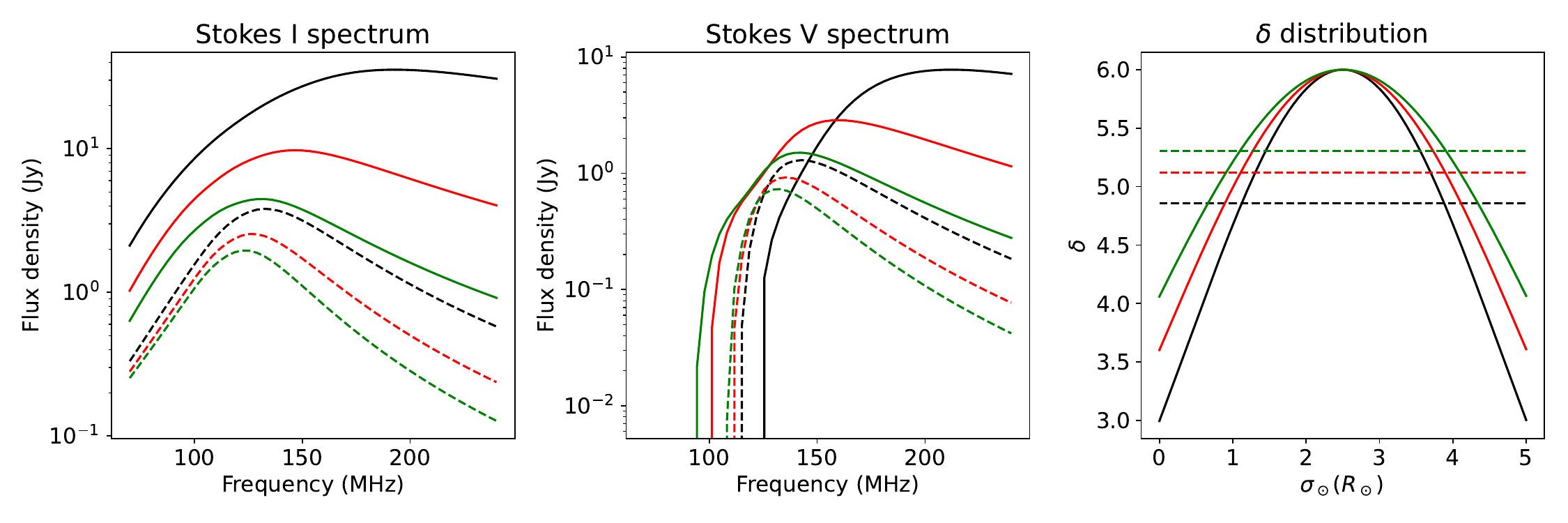}
    \caption{\textbf{Simulated GS spectra for inhomogeneous and mean homogeneous distribution of CME plasma parameters. First column: }Simulated Stokes I spectra. \textbf{Second column:} Simulated Stokes V spectra. \textbf{Third column: }Distribution of plasma parameters along the LoS. Simulated spectra for inhomogeneous and mean homogeneous distributions are shown by solid and dashed lines, respectively.}
    \label{fig:simulation_2_1}
\end{figure*}

\subsection{Insufficiency of Homogeneous GS Model}\label{subsec:obs_evidence_inhom}
K23 demonstrated that including even the sensitive upper limits from Stokes V observations along with the Stokes I spectrum can significantly improve the ability to constrain the GS model parameters. The preceding discussion shows that despite the availability of stringent Stokes V upper limits and Stokes I spectrum, a Stokes V detection even at a single spectral point can provide significant additional information. However, the routinely used GS modeling approach is unable to find a model consistent with Stokes V detection, Stokes V upper limits, and Stokes I spectrum simultaneously. This led us to critically examine the assumptions made during the GS modeling process. We have just demonstrated that considering more general electron energy distributions and accounting for the impact of their pitch-angle distributions, while holding the assumption of homogeneity, are not sufficient to meet these constraints. The only remaining assumption is that of homogeneity in the CME plasma, both along the LoS and within the PSF-sized region, being modeled. Given that it is already well known that the plasma and magnetic field comprising a CME are inhomogeneous, the need for exploring such models is hardly surprising. On the other hand, the Stokes I spectra have always been found to be consistent with homogeneous and isotropic GS models in this and earlier studies. Since all earlier studies only used the Stokes I spectrum to model the GS emission, there was never any need to consider an inhomogeneous GS model.

Given the Stokes V detection only at a single frequency, it is not possible to constrain an inhomogeneous GS model using the current observations. But, in light of the demonstration of the inability of the homogeneous models to explain both Stokes I and V observations simultaneously, we take the first steps toward understanding the impact of inhomogeneity on the GS spectra using simple toy simulations. The results from these toy simulations will help assess the sensitivity of different GS model parameters to inhomogeneity. As and when Stokes V measurements become available across the band this information will serve to guide the modeling process. These toy simulations are described next.

\section{Effects of Inhomogeneity on GS Spectrum}\label{sec:nonuniform}
A homogeneous and isotropic GS model has ten free parameters. For an inhomogeneous GS model, the number of free parameters will increase substantially. Modeling these large numbers of free parameters with a limited number of spectral measurements reduces the ability to effectively constrain them. Hence, we start by trying to isolate and quantify the impact of inhomogeneity in individual parameters of the GS models on the Stokes I and V GS spectra using toy models, so that the parameter with higher sensitivities to inhomogeneity can be identified.

\subsection{Description of the Simulation}\label{subsec:simulation}
We have performed two simulations to understand the effects of inhomogeneities on the GS spectra. Two of the ten GS model parameters are related to the geometry of the CME ($A$ and $L$) and the concept of inhomogeneity does not apply to them. For all other parameters, except $\theta$, we have simulated the Stokes I and V spectra for a Gaussian distribution of these parameters along the LoS.

\subsubsection{Simulation 1: Effects of Gaussian Distribution of Plasma Parameters}\label{subsec:inhomo_smooth}
Stokes I and V spectra are simulated for different plasma parameters following a Gaussian distribution along the LoS having the form
\begin{equation}
    p(l)=p_0\ exp\left[-\left(\frac{l-\frac{l_\mathrm{max}}{2}}{\sigma}\right)^2\right],
    \label{eq:inhomo_dist}
\end{equation}
where $p(l)$ is the value of the plasma parameter at LoS segment $l$, $p_0$ represents the maximum value of the plasma parameter, $l_\mathrm{max}$ is the maximum length of the LoS and $\sigma$ the width of the distribution. $\sigma$ values are presented in units of $R_\odot$ in Figure \ref{fig:simulation_1} and are denoted by $\sigma_\odot$. For each of the parameters, a fiducial value is identified within a physically motivated range of values and is used as the maximum value ($p_0$) for the simulation. These fiducial values are -- i) $|B|_0=6\ \mathrm{G}$, ii) $\theta=80^{\circ}$, iii) $A = 10^{20}\ \mathrm{cm^{2}}$, iv) $T_0= 10^6\ \mathrm{K}$, v) $n_\mathrm{thermal,0}= 1.25\times10^6\ \mathrm{cm^{-3}}$, vi) $n_\mathrm{nonth,0}= 1.25\times10^4\ \mathrm{cm^{-3}}$, vii) $\delta= 4$, viii) $L= 5\times10^{10}\ \mathrm{cm}$, ix) $E_\mathrm{min}=10\ \mathrm{keV}$ and x) $E_\mathrm{max}= 15\ \mathrm{MeV}$. We divided the GS source into 1000 LoS segments for this simulation, each of length 0.005 $R_\odot$. $\sigma$ is varied between 1 $R_\odot$ and 5 $R_\odot$. These limits are chosen in such a way that at the lowest $\sigma$ the Gaussian distribution is highly peaked and essentially vanishes beyond a few solar radii, while at the highest $\sigma$ the distribution comes close to the usual homogeneous distribution considered in earlier simulations. These choices have been made to allow us to examine the impact of homogeneity on the spectra.

In these toy simulations, only one GS plasma parameter is considered to be inhomogeneous at any time. All other model parameters are regarded to be homogeneous and set to their respective fiducial values. Simulated Stokes I and V spectra are shown in Figures \ref{fig:simulation_1}. It is evident from the bottom row of Figure \ref{fig:simulation_1} that the inhomogeneity in $n_\mathrm{thermal}$ and $T$ does not have any effect on the observed Stokes I or V spectra. Other plasma parameters -- $|B|,\ \delta,\ n_\mathrm{nonth}$ and $E_\mathrm{min}$ -- show significant effects on both Stokes I and V spectra. With increasing inhomogeneity (i.e., decreasing $\sigma$), the peak frequency and peak flux density of the Stokes I and V spectra decreases for $n_\mathrm{nonth}$, $E_\mathrm{min}$ and $|B|$. It is also found that the effect of inhomogeneity on the Stokes I spectra is seen only in the optically thin part of the spectra (i.e. at frequencies higher than the peak of the spectrum), while it is evident in both optically thick and thin parts of the Stokes V spectra. For both Stokes I and V, the flux density decreases with increasing inhomogeneity in $|B|, n_\mathrm{nonth}$ and $E_\mathrm{min}$, and the spectral peak shifts to lower frequencies. Unlike these three parameters, even a low level of inhomogeneity in $\delta$ leads to significant changes in both Stokes I and V spectra.

\subsubsection{Simulation 2: Effects on Modeling an Inhomogeneous GS Source with Homogeneous Model}
\label{subsec:simulation_2}
Since homogeneous GS models have routinely been used to model GS emissions, it is also important to understand their efficacy at estimating the mean value of the distribution of the relevant plasma parameters. To understand this, we first simulated the Stokes I and V GS spectra corresponding to a Gaussian distribution of the model parameters along the LoS, taken one at a time -- $|B|$, $n_\mathrm{nonth}$, $E_\mathrm{min}$ and $\delta$ -- for a few different Gaussian widths. All other parameters were fixed at the fiducial values mentioned in Section \ref{subsec:inhomo_smooth}. The corresponding GS spectra are shown by solid lines in Figure \ref{fig:simulation_2_1}. We then compute the GS spectra corresponding to a homogeneous model with the Gaussian distribution replaced by its mean value. The mean values are marked by dashed lines in Figure \ref{fig:simulation_2_1} and the corresponding simulated Stokes I and V spectra by dashed lines in the same figure. 

It is evident from the first row of Figure \ref{fig:simulation_2_1} that while different Gaussian distributions of $n_\mathrm{nonth}$ produce different Stokes I and V spectra, the corresponding homogeneous GS model set to the mean value of $n_\mathrm{nonth}$ lead to essentially identical Stokes I and V spectra. This implies that a homogeneous GS model can provide an accurate estimation of the mean $n_\mathrm{nonth}$ along the LoS. For the other three parameters -- $|B|$, $E_\mathrm{min}$ and $\delta$ -- there are significant differences between spectra resulting from inhomogeneous and homogeneous distributions set to the corresponding mean value of the parameter, inhomogeneity in which is being explored. These differences grow larger as the width of Gaussian distribution grows smaller, i.e. with an increase in the degree of inhomogeneity along the LoS. Conversely, as should be expected, the two spectra come closer to each other as $\sigma$ increases and the degree of inhomogeneity decreases.

These toy simulations show that the ability of a homogeneous model to represent the mean of the true inhomogeneous distribution for $|B|$, $E_\mathrm{min}$ and $\delta$, even for the simplest of inhomogeneous models, is dependent on the level of inhomogeneity present in the medium and grows poorer with increasing degree of inhomogeneity. For large enough levels of inhomogeneity, the GS model spectra will differ significantly from the ones corresponding to the mean values of the distributions.
\begin{table*}[!htbp]
    \centering
    \renewcommand{\arraystretch}{1.5}
    \begin{tabular}{|p{1.6cm}|p{5cm}|p{5cm}|p{5cm}|} \hline 
         GS Model Parameter&  Stokes I Spectrum &  Stokes V Spectrum&  Effects of Using Homogeneous Model\\ \hline \hline 
         $n_\mathrm{nonth}$ & Peak flux density decreases with increasing inhomogeneity. Only optically thick part of the spectrum is affected. & Peak flux density decreases with increasing inhomogeneity. Both optically thin and thick part of the spectrum are affected. &  Mean value of $n_\mathrm{nonth}$ is recovered.\\ \hline 
         $|B|$, $E_\mathrm{min}$ & Peak flux density decreases with increasing inhomogeneity. Only optically thick part of the spectrum is affected. & Peak flux density decreases with increasing inhomogeneity. Both optically thin and thick part of the spectrum are affected. &  Do not recover mean value of the distribution. \\ \hline 
         $\delta$ & Similar changes observed with smaller inhomogeneity. & Similar changes observed with smaller inhomogeneity. &   Do not recover mean value of the distribution. \\ \hline
         $T$, $n_\mathrm{thermal}$ & No effect & No effect & No effect \\ \hline
    \end{tabular}
\caption{Effects of inhomogeneity of GS model parameters}
\label{tab:inhomo_effect_table}
\end{table*}

\section{Conclusion}\label{sec:conclusion}
As pointed out in Section \ref{sec:intro}, detection of CME GS emission is challenging and this has resulted in only a handful of successful studies being available in the literature. The dataset chosen for this study was particularly challenging. Radio emission from this event is faint and the observation was done at the lowest allowed elevation of the MWA, where the instrument has its lowest sensitivity. Stokes V flux density being only a fraction of Stokes I, implies that Stokes V detection is even more challenging than that for Stokes I. Nonetheless, this study has detected Stokes I emission with high significance over a spatially extended region and across essentially the entire MWA observing band. We have also detected Stokes V emission, though at a single spectral point and over a much smaller spatial extent, and placed sensitive upper limits on fractional circular polarization over the rest of the region where Stokes I emission was detected.

For the first time, this work uses Stokes V detection and upper limits jointly with the Stokes I spectrum to constrain the GS model parameters of a CME. A similar approach, to model GS emission from stellar emissions using both Stokes I and V spectra, has also been attempted recently \citep{Golay2023}. As additional observational constraints become available, one expects to be able to better constrain the GS models. Contrary to this, we find that no model in the reasonable part of the solution phase space can meet all of the observational constraints imposed by the data. This situation, where less constraining data leads to a good model fit and more constraining data does not, strongly suggests that one should examine the model and the assumptions it makes. It is well known from magnetic flux-rope models \citep[e.g.,][etc.]{Isavnin_2016,Mostl2018} that CME plasma parameters are not expected to be homogeneous along the LoS. Nonetheless, they have always been assumed to be so in all prior works studying CME GS emission. In fact, these simplifying assumptions have been essential because the constraints available from the observations are not enough to constrain the more detailed and physically meaningful GS source models which require many more free parameters. Among other things, this work presents a systematic, though limited, study of the impact of the violation of the assumption homogeneity on the observed GS spectra.

To explore the reasons behind the inability of GS models used to describe the observed spectra, we systematically explored the possibility of the assumptions made by the GS models used being violated -- electron energy distribution being different from a single power law; anisotropy in electron pitch-angle distribution; and inhomogeneities in the distribution of any of the many plasma parameters along the LoS. In the limited but illuminating exploration done here, it is found that a homogeneous GS model with different electron energy distributions or different pitch-angle distributions cannot reproduce the observed Stokes I and V simultaneously. To build a quantitative sense for the impact of inhomogeneity on the observed Stokes I and V spectra, we carried out a set of toy simulations where only one of the GS model parameters was allowed to vary along the LoS in a Gaussian manner, while all others were held constant, i.e. deemed to be homogeneous. These toy simulations allowed us to identify the effects of inhomogeneity in different plasma parameters individually on Stokes I and V GS spectra. They led to the conclusion that inhomogeneity in thermal electron distribution and temperature have negligible impact on the observed GS spectra. Inhomogeneity in magnetic field and nonthermal electrons, on the other hand, were found to lead to significant effects on both the Stokes I and V GS spectra. 

All of the GS observations have, thus far, been modeled assuming the plasma filling the modeled volume to be homogeneous. Hence, in light of the above evidence for the insufficiency of homogeneous GS models, it is useful to ask the following question -- if the observed GS spectra are truly arising from inhomogeneous plasma distributions, how close these spectra are to those generated using the mean value of this distribution? A corollary is that if we were to fit these spectra using homogeneous GS models, how close would the best fit GS parameter values be to the means of the distribution? We carried out a set of simplistic simulations to examine this using toy models where GS model parameters were distributed along the LoS in an inhomogeneous manner only one at a time and compared with model spectra generated using the mean of the distribution. These led to the conclusion that inhomogeneities in $n_\mathrm{nonth}$ have very little impact on the observed GS Stokes I and V spectra. However, for inhomogeneities in distributions of $|B|$, $E_\mathrm{min}$, and $\delta$, the similarity of the true inhomogeneous spectra to those generated using mean homogeneous GS models depends on the level of inhomogeneity. As the distribution of plasma parameters becomes increasingly inhomogeneous, the difference between the two grows larger and the GS spectra cannot be correctly modeled assuming homogeneous models. These findings are summarized in Table \ref{tab:inhomo_effect_table}.

The CME studied here is quite weak and the site of GS emission lies well behind the CME leading edge. It coincides with the location of a streamer which also shows some signatures of interaction with the CME. Hence, it seems very likely that the majority of the nonthermal electrons giving rise to GS emission originate from the small-scale magnetic reconnections between the CME and the streamer magnetic fields. Possible reasons that might lead to inhomogeneities in the volume modeled for GS emissions are -- the nonthermal electrons are yet to homogenize and isotropize and the presence of reconnection sites themselves introduces significant inhomogeneity. Perhaps this forms an important distinguishing factor between this and other instances where GS emissions from CMEs have been studied.

\section{Future Work}\label{sec:future_work}
On the observational front, we note that the noise levels of the solar images achieved by P-AIRCARS used here are usually within a factor of $\sim1.5$ of the expected thermal noise \citep{Kansabanik2022}. This implies that we are close to the limits of what can be delivered by these data. So more sensitive observations, needed to routinely detect/place stronger upper limits on Stokes V over larger volumes, will need to wait for future more sensitive instruments. The MWA Phase-III is already being pursued and will double its collecting area. We are actively working on enabling solar observations with the mid-frequency precursor of the Square Kilometre Array Observatory (SKAO), MeerKAT \citep{Kansabanik2023_meerkat}. This will be the best available instrument for pushing this science to higher frequencies. Hence it will allow us to perform similar studies at lower heliospheric distances and will also provide higher angular resolution. Both the low- and mid-frequency telescopes of the SKAO are expected to become available later this decade and will enable observations with much higher sensitivity and imaging fidelity.

On the modeling front, it is reasonable to expect the reality to only be more complex, with the distributions of multiple, perhaps all, parameters being inhomogeneous and more complicated than the simplest possibility of a Gaussian distribution considered for our toy simulations. The currently available data and/or modeling approaches do not have the ability to constrain the much larger number of free parameters required for an inhomogeneous model. However, one aspect that is yet to be explored in the GS modeling of CMEs is to model the entire spatially resolved emission as coming from a single large structure, rather than the current practice of treating each PSF-sized region independently. By ignoring the physical continuities that must exist in the distribution of plasma parameters this approach, on the one hand, artificially inflates the number of model parameters that need to be constrained and, on the other, makes sub-optimal use of the available constraints.

As our ability to detect spatially resolved Stokes I and V emission over extended regions and the angular resolution of our images improves, this approach will become increasingly interesting to explore. By trying to constrain a single spatially varying model spanning the entire extent of the detected emission, this approach has the potential to significantly reduce the number of free parameters in the problem. Though, not an apples-to-apples comparison, similar physics-based three-dimensional modeling of GS emission from flare loops have already been performed \citep[e.g.][]{Kuznetsov_2011,Reznikova_2014,Doorsselaere2016}. Using spectroscopic imaging observations of flares at microwave frequencies, these models have been used to constrain flare parameters. Following a similar approach for constraining GS emission from CMEs can be the next step in this area of research.

We believe that with the confluence of availability of more sensitive images spanning a broader spectral coverage and with higher angular resolution expected to become available from upcoming instruments; and the insights gained here, we are well poised to place meaningful constraints on CME physical parameters using their GS emission.

\facilities{Murchison Widefield Array \citep[MWA;][]{lonsdale2009,Tingay2013},Solar and Heliospheric Observatory \citep[SOHO;][]{Domingo1995}, Solar Terrestrial Relations Observatory \citep[STEREO,][]{Kaiser2008}}

\software{astropy \citep{astropy:2013,astropy:2018,astropy:2022}, matplotlib \citep{Hunter:2007}, Numpy \citep{Harris2020}, CASA \citep{mcmullin2007,CASA2022}, P-AIRCARS \citep{paircars_zenodo}, GCS-python \citep{gcs_python}, GScode \citep{GS_code2021}, JHelioviewer \citep{JHelioviewer}}\\

\noindent This scientific work makes use of the Murchison Radio-astronomy Observatory (MRO), operated by the Commonwealth Scientific and Industrial Research Organisation (CSIRO). We acknowledge the Wajarri Yamatji people as the traditional owners of the Observatory site.  Support for the operation of the MWA is provided by the Australian Government's National Collaborative Research Infrastructure Strategy (NCRIS), under a contract to Curtin University administered by Astronomy Australia Limited. We acknowledge the Pawsey Supercomputing Centre, which is supported by the Western Australian and Australian Governments. D.K. gratefully acknowledges Barun Maity (Max Plank Institute of Astronomy) for useful discussions. We also thank the anonymous referee for the comments and suggestions, which have helped improve the clarity and presentation of this work. D.K. acknowledges the support by the NASA Living with a Star Jack Eddy Postdoctoral Fellowship Program, administered by UCAR’s Cooperative Programs for the Advancement of Earth System Science (CPAESS) under award \#80NSSC22M0097. D.K. and D.O. acknowledge the support of the Department of Atomic Energy, Government of India, under project no. 12-R\&D-TFR-5.02-0700.

\bibliography{sample631}{}
\bibliographystyle{aasjournal}

\end{document}